\documentclass[12pt,a4paper]{article}
\pdfoutput=1

\usepackage{multirow}
\usepackage{amsmath}
\usepackage{color}
\usepackage{amsfonts}
\usepackage{amssymb}
\usepackage{graphicx}
\usepackage{geometry}
\usepackage{amssymb,epsfig,subfigure}
\usepackage{hyperref}
\usepackage{url}
\usepackage{comment}
\usepackage[font=footnotesize]{caption}
\usepackage{slashed}

\makeatletter
\renewcommand\section{\@startsection {section}{1}{\z@}%
                                 {-3.5ex \@plus -1ex \@minus -.2ex}%nn
                                   {2.3ex \@plus.2ex}%
                                   {\normalfont\large\bfseries}}
\renewcommand\subsection{\@startsection{subsection}{2}{\z@}%
                                   {-3.25ex\@plus -1ex \@minus -.2ex}%
                                     {1.5ex \@plus .2ex}%
                                     {\normalfont\bfseries}}
\renewcommand\subsubsection{\@startsection{subsubsection}{3}{\z@}%
                                   {-3.25ex\@plus -1ex \@minus -.2ex}%
                                     {1.5ex \@plus .2ex}%
                                     {\normalfont\itshape}}
\makeatother

\def\pplogo{\vbox{\kern-\headheight\kern -29pt
\halign{##&##\hfil\cr&{\ppnumber}\cr\rule{0pt}{2.5ex}&\ppdate\cr}}}
\makeatletter
\def\ps@firstpage{\ps@empty \def\@oddhead{\hss\pplogo}%
  \let\@evenhead\@oddhead % in case an article starts on a left-hand page
}%      The only change in \maketitle is \thispagestyle{firstpage} instead of 
\def\maketitle{\par
 \begingroup
 \def\thefootnote{\fnsymbol{footnote}}
 \def\@makefnmark{\hbox{$^{\@thefnmark}$\hss}}
 \if@twocolumn
 \twocolumn[\@maketitle]
 \else \newpage
 \global\@topnum\z@ \@maketitle \fi\thispagestyle{firstpage}\@thanks
 \endgroup
 \setcounter{footnote}{0}
 \let\maketitle\relax
 \let\@maketitle\relax
 \gdef\@thanks{}\gdef\@author{}\gdef\@title{}\let\thanks\relax}
\makeatother

\numberwithin{equation}{section}

\newcommand\eea{\end{eqnarray}}
\newcommand\bea{\begin{eqnarray}}

\def\beq{\begin{equation}}
\def\eeq{\end{equation}}

\newcommand{\be}{\begin{equation}}
\newcommand{\ee}{\end{equation}}
\newcommand{\ba}{\begin{align}}
\newcommand{\ea}{\end{align}}
\newcommand{\bg}{\begin{gather}}
\newcommand{\eg}{\end{gather}}
\newcommand{\bseq}{\begin{subequations}}
\newcommand{\eseq}{\end{subequations}}

\renewcommand{\tanh}{\mathop{\rm th}\nolimits}

\renewcommand{\t}{\tilde}

\newcommand{\mc}{\mathcal}

\newcommand{\dd}[1]{\text{d}#1}
\newcommand{\of}[1]{\left(#1\right)}
\newcommand{\off}[1]{\left[#1\right]}

\newcommand{\coment}[1]{}

\begin{document}
\setcounter{page}0
\def\ppnumber{\vbox{\baselineskip14pt
%\hbox{hep-th/0000000}
}}
\def\ppdate{
%\footnotesize{SU/ITP-14/XX}
} \date{}

\author{Nicolás Abate$^{1, 2 }$, Ignacio Salazar$^{3}$, Gonzalo Torroba$^{1, 2}$\\
[7mm] \\
{\normalsize \it $^1$Centro At\'omico Bariloche and CONICET}\\
{\normalsize \it $^2$ Instituto Balseiro, UNCuyo and CNEA}\\
{\normalsize \it S.C. de Bariloche, R\'io Negro, R8402AGP, Argentina}\\
{\normalsize \it $^3$ Instituto de F\'\i sica de La Plata - CONICET, C.C. 67, 1900 La Plata, Argentina}\\
}

\bigskip
\title{\bf  Entanglement and Renormalization Group Irreversibility of Quantum Field Theory \\ in AdS
\vskip 0.5cm}
\maketitle

\begin{abstract}
We study nonperturbative aspects of quantum field theory (QFT) in rigid anti–de Sitter (AdS) spacetime using quantum information–theoretic methods. While irreversibility of renormalization group (RG) flows is well established in flat space, it is not obvious whether it persists in AdS, where negative curvature and an asymptotic time-like boundary significantly modify infrared dynamics.
Using strong subadditivity and AdS invariance, we derive an entropic second-order differential inequality for the difference between the vacuum entanglement entropy of a QFT and that of its ultraviolet fixed point, evaluated for spherical bulk regions. This inequality allows us to define RG charges that measure the relevant number of degrees of freedom, and we prove the irreversibility of the RG in $2,\,3,$ and $4$ spacetime dimensions.
We further analyze free scalar and fermion theories in AdS, developing lattice formulations adapted to the geometry and computing entanglement entropies and RG charges. In AdS$_2$, we obtain analytic results for a massive Dirac fermion and compare them with numerical lattice calculations. These examples illustrate the general irreversibility theorem and clarify the distinction between conformal and massive theories in AdS.
\end{abstract}
\bigskip

\newpage

\tableofcontents

\vskip 1cm

%%%%%%%%%%%%%%%%%%%%%%%%%%%%%%%%%%%%%%%%%
%%%%%%%%%%%%%%%%%%%%%%%%%%%%%%%%%%%%%%%%%
%%%%%%%%%%%%%%%%%%%%%%%%%%%%%%%%%%%%%%%%%
%%%%%%%%%%%%%%%%%%%%%%%%%%%%%%%%%%%%%%%%%
\section{Introduction}\label{sec:intro}

Understanding quantum field theory (QFT) in a rigid anti–de Sitter (AdS) spacetime is important both conceptually and technically. On the one hand, QFT in AdS underlies many aspects of the AdS/CFT correspondence, providing a controlled setting to analyze bulk dynamics and quantum effects beyond semiclassical gravity (see e.g. \cite{Duetsch:2002hc, Faulkner:2013ana, Engelhardt:2014gca, Benedetti:2022aiw}). It also plays a central role in modern bootstrap approaches formulated in AdS (see \cite{Paulos:2016fap, Komatsu:2020sag, vanRees:2022zmr}) and in the study of quantum field theories with boundaries and defects (see e.g. \cite{Giombi:2020rmc, Bason:2025sxb}). On the other hand, even independently of holography, QFT in rigid AdS defines a distinctive dynamical framework: the presence of negative curvature and an asymptotic time-like boundary qualitatively modifies infrared behavior, operator correlators, and the structure of renormalization group flows.

The negative curvature leads to exponential growth of volume with distance,  mimicking certain features of infinite-dimensional geometries, such as exponential volume growth and correlator suppression. On the other hand, the presence of the boundary allows to probe the bulk dynamics in ways that are not available in flat spacetime, by changing boundary conditions and inserting boundary sources. At the same time, AdS is maximally symmetric and this provides an enhanced control over correlation functions and, as we shall see, entanglement properties. An early foundational work on QFT in AdS which emphasized some of these aspects is \cite{Callan:1989em}. Conformal field theories (CFTs) in rigid AdS were studied in \cite{Aharony:2010ay}, including free theories and their allowed boundary conditions. AdS has been very useful for understanding confinement \cite{Aharony:2012jf, Ciccone:2024guw, Ghodsi:2024jxe, Ciccone:2025dqx}; different renormalization group flows in AdS are also being actively analyzed, exhibiting rich phenomena absent from flat space \cite{Carmi:2018qzm, Lauria:2023uca, Copetti:2023sya}. Unlike flat space, in AdS the conformal group is not enhanced beyond the isometry group; and related to this, the exponential growth of the spacetime gives rise to exponentially suppressed correlators at long distance, making nontrivial the distinction between conformal and massive theories.

In this work we study nonperturbative properties of QFT in AdS using methods from quantum information theory. This approach has been extremely fruitful in flat space for establishing the irreversibility of renormalization group (RG) flows \cite{Casini:2004bw, Casini:2012ei, Casini:2017vbe}, in the context of theories with defects and boundaries \cite{Casini:2016fgb, casini2019irreversibility, casini2023entropic, Casini:2023kyj}, and more recently for QFT in de Sitter \cite{Abate:2024xyb, Abate:2024nyh}. We consider $d$-dimensional QFTs that are unitary and AdS invariant, and which can be obtained by perturbing CFTs by relevant interactions. This produces RG flows which cover a space of QFTs with well-defined UV completions. Analyzing the properties of this space of QFTs is much more nontrivial than in flat space due to the presence of the boundary, the interplay of the AdS radius with the RG energy scale, and the more subtle infrared dynamics of massive theories. Because of these new aspects in AdS, it is not clear whether the irreversibility of the RG also holds. One of our main results will be that it does.

Using strong subadditivity and AdS invariance, we derive the following second order differential inequality for $\Delta S(R)=S_{\mathrm{QFT}}(R)- S_{\mathrm{UV}}(R)$, 
\be\label{eq:introDS}
R\, \Delta S''(R) -(d-3) \Delta S'(R) \le 0\,,
\ee
where we consider spherical entangling regions inside AdS with area proportional to $R^{d-2}$, $S_{\mathrm{QFT}}(R)$ is the vacuum entanglement entropy for the QFT with nontrivial RG flow, and $S_{\mathrm{UV}}(R)$ is the entanglement entropy for the UV fixed point.\footnote{We stress that the entangling regions considered in this work are bulk regions not anchored to the boundary, as opposed to the ones that appear naturally for holographic Ryu-Takanagi calculations \cite{Ryu:2006bv}. The entanglement entropy here is computed entirely within the setup of QFT in rigid AdS, without gravitational effects.} This inequality was first proved in flat space in \cite{Casini:2017vbe}, and more recently in de Sitter by \cite{Abate:2024nyh}. With this result of the present paper, the inequality is then demonstrated to be valid for all maximally symmetric spacetimes in $d$-dimensions. Even though the inequality is the same in the three spacetimes (Minkowski, dS and AdS) when expressed in terms of the entangling area, the physical consequences are qualitatively different, as seen from the fact that the relation between the area and the proper radius depends crucially on the curvature of the space-time. In AdS, at long distance $R$ grows exponentially with the proper radius of the entangling region.

 The inequality leads to two immediate consequences. First, it will allow us to define an RG charge in AdS that measures the relevant number of degrees of freedom. At short distances this reduces to the standard flat space RG charges (C, F, or A), while for distances larger than the AdS scale it is a genuinely new curved spacetime quantity. This is particularly helpful for the problem of distinguishing conformal and massive theories in AdS, which, as we discuss below, is much more nontrivial than in flat space. Secondly, it implies that RG flows in rigid AdS are irreversible in $d=2, 3$ and $4$ space-time dimensions. Therefore, even though the asymptotic time-like boundary of AdS and its negative curvature have strong effects on the dynamics, they do not spoil the monotonicity of the RG, measured in terms of the proper notion of RG charge that we will construct. In analogy to flat space, we will refer to these as the C, F and A theorems in AdS.

The second goal of this work is to study entanglement properties of free QFTs in AdS. Based on lessons from this analysis in flat space (reviewed e.g. in \cite{Casini:2009sr}), we expect that this effort will shed light on more general aspects of interacting models. Here we focus on RG flows for free scalar and fermion fields, and it would be interesting to extend the analysis to gauge fields and gravitons. The main tool will be the development of lattice calculations in AdS, a topic which only recently has begun to be explored in the literature \cite{Boutivas:2025ksp, Govindarajan:2026psk}. We will address new challenges that spatial lattices in AdS bring forward, including the emergence of AdS isometries in the continuum limit, and the existence of a Markovian cutoff. These results will also provide explicit examples of RG charges and the general irreversibility result.

The paper is organized as follows. In Sec. \ref{sec:qftads} we present a review of QFT in AdS, focusing on CFTs and RG flows. In Sec. \ref{sec:markov} we analyze the entanglement structure of CFTs in AdS. This requires considering entangling regions whose boundaries are general surfaces in a common light-cone. Null variations of these entangling boundaries are extremely important, as they will allow us to establish the validity of the Markov property for CFTs in AdS. From here we will determine the general form of the entanglement entropy (EE) at fixed points. This includes area and subleading terms, as well as universal RG charges, related by a Weyl transformation to the flat space result. In Sec. \ref{sec:irrev} we consider general QFTs in AdS, obtained as RG flows of UV CFTs, and prove the irreversibility formula (\ref{eq:introDS}); we discuss its implications for the definition of RG charges and the irreversibility of the RG in AdS. The last two sections are devoted to free field examples. In Sec. \ref{sec:fermion} we consider a massive Dirac fermion in AdS$_2$. We derive analytic results for the entanglement entropy and running C-function using bosonization and the solution to the Painlevé VI equation. We also consider a lattice fermion, evaluate numerically the entropy and the C-function, and compare with the analytic predictions (finding excellent agreement). This 2d model serves as a prototype for understanding and solving lattice issues in AdS. In Sec. \ref{sec:latticed} we consider lattice models for massive scalars in $d=2,\,3$ and $4$ space-time dimensions, studying their respective RG charges and evaluating (\ref{eq:introDS}). Finally, we present our concluding remarks and future directions in Sec. \ref{sec:concl}.

%%%%%%%%%%%%%%%%%%%%%%%%%%%%%%%%%%%%%%%%%
%%%%%%%%%%%%%%%%%%%%%%%%%%%%%%%%%%%%%%%%%
%%%%%%%%%%%%%%%%%%%%%%%%%%%%%%%%%%%%%%%%%
%%%%%%%%%%%%%%%%%%%%%%%%%%%%%%%%%%%%%%%%%
\section{Quantum field theory in AdS}\label{sec:qftads}

In this section we begin by discussing some properties of QFTs in AdS that will be needed for our purpose.

\subsection{Review of AdS geometry}\label{subsec:geomAdS}

\begin{figure}[h]
    \centering
    \includegraphics[width=0.35\textwidth]{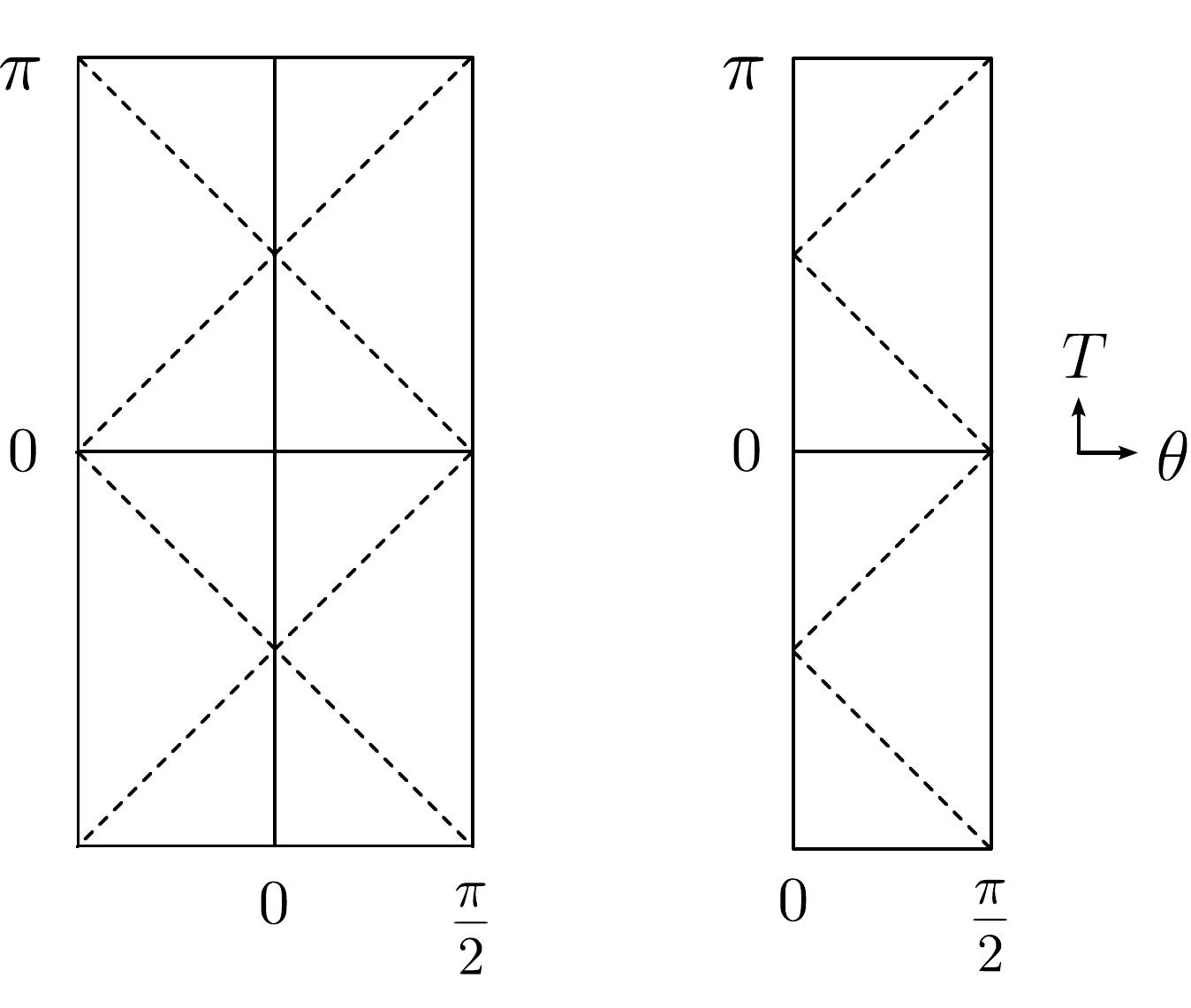}
    \caption{Penrose diagram for AdS$_d$ for $d=2$ (left) and $d>2$ (right).}
    \label{fig:pd}
\end{figure}

Recall that $d$ dimensional Anti-de Sitter space-time (AdS$_d$) is defined as an embedded hyperboloid in $\mathbb
R^{d-1,2}$
\be
-(X^0)^2+(X^i)^2-(X^d)^2=-\ell^2\ ,\quad i=1,\dots,d-1\,,
\ee
where $\ell$ is the AdS radius. Choosing the parametrization
\be
X^0=\ell\,\frac{\cos T}{\cos\theta}\,,\, X^i=\ell\,\tan\theta\,\hat{x}^i\,,\, X^d=\ell\,\frac{\sin T}{\cos\theta}\,,
\ee
with $0\leq T\leq2\pi$,\footnote{One can then take the global covering $T\in\mathbb{R}$ in order to avoid closed time-like curves} and $0\leq\theta\leq\pi/2$ (or $|\theta|\leq\pi/2$ in $d=2$), gives the following metric on AdS$_d$:
\be\label{eq:metricAdS}
\dd s^2_{AdS}=\frac{\ell^2}{\cos^2\theta}\of{-\dd T^2+\dd\theta^2+\sin^2\theta\,\dd\Omega_{d-2}^2}.
\ee
In these coordinates, which cover the whole manifold, the light rays travel at $45^{\circ}$ and thus they define the Penrose diagram of AdS$_d$, which we show in Fig. \ref{fig:pd}.

We will also find it useful to work in global AdS in terms of the proper distance at fixed time $\rho$. Integrating $d\theta/\cos \theta= d\rho$ gives
\be
\rho={\textrm {arcsinh}}(\tan \theta)
\ee
and the metric becomes
\be
\label{eq:global_ads}
\dd s^2_{AdS}=\ell^2 \left(- \cosh^2 \rho\, \dd T^2+ \dd \rho^2+ \sinh^2 \rho\,\dd\Omega_{d-2}^2\right)\,.
\ee

\begin{comment}
Another possibility is to use Poincaré coordinates:
\be
X^0=\frac{z^2-t^2+\delta^{ij}x_i x_j+\ell^2}{2z},\ X^i=\frac{\ell x^i}{z},\ X^{d-1}=\frac{z^2+t^2-\delta^{ij}x_i x_j-\ell^2}{2z},\ X^d=\frac{\ell t}{z}\,,
\ee
where $i=1,\dots,d-2$ and $z>0$. This covers half of the spacetime -- the so-called Poincaré Patch. One of the boundaries of AdS happens at $z=0$. The metric in these coordinates is conformally flat
\be
\dd s^2=\frac{\eta_{\mu\nu}\dd x^\mu \dd x^\nu }{\ell^2 z^2}\,,
\ee
for $x^\mu=(t,x^i,z)$.
\end{comment}

The isometries are given by the Lorentz transformations in the embedding which preserve the hyperboloid. This is the group $SO(d-1,2)$, generated by $J_{MN}=X_M\partial_N-X_N\partial_M$. Besides the $S^{d-2}$ rotations $J_{ij}$, the isometry generators are the global time translation $J_{0d}$ and the ``boosts'' $J_{0i}$ and $J_{di}$. 

In any coordinate system, the proper length can be computed using the metric, obtaining
\be\label{eq:L}
L=\ell\,\mathrm{arccosh}\,\sigma\,,
\ee
where $\sigma=X_1\cdot X_2$ is the dot product between the endpoints of the geodesic in the embedding space. In global conformal coordinates, this is
\be
\sigma=\tan\theta_1\tan \theta_2 -\cos(T_1-T_2)\sec\theta_1\sec\theta_2\,.
\ee
%{\color{blue} GT: me estaba quedando un menos global acá}

%while in Poincaré coordinates
%\be
%\sigma=\frac{z_1^2+z_2^2+(t_1-t_2)^2}{2z_1 z_2}\,.
%\ee

\subsection{CFTs and renormalization group flows in AdS}

Lorentzian CFTs are naturally defined on the Lorentzian cylinder (LC) $\mathbb R \times S^{d-1}$ (see \cite{Mack:1988nf} for a nice exposition),
\be\label{eq:dsLC}
\dd s^2_{LC}=\ell^2\of{-\dd T^2+\dd\theta^2+\sin^2\theta\,\dd\Omega_{d-2}^2}\,,
\ee
where $0\le \theta \le \pi$, and $\ell$ is the radius of the cylinder. The conformal group in the LC is $SO(d,2)$. In the euclidean theory, time evolution along the cylinder is conformally equivalent to radial evolution, and the energies of primary states on $S^{d-1}$ map to the conformal dimensions.

Minkowski space-time $\mathbb R^{d-1,1}$ is conformally equivalent to the LC (see e.g. \cite{Candelas:1978gf}). To see this, we define null coordinates
\be
\theta_\pm = \theta \pm T\;,\; r_\pm= r \pm t\,,
\ee
and perform the coordinate transformation 
\be\label{eq:rpmdef}
r_\pm = \ell \tan \left(\frac{\theta_\pm}{2}\right)\,.
\ee
The metric then becomes
\be\label{eq:LCMink}
\dd s^2_{LC}= \frac{4}{(1+(r_+/\ell)^2)(1+(r_-/\ell)^2)}\,\dd s^2_\textrm{Mink}
\ee
with
\be
\dd s^2_\textrm{Mink}= -dt^2+ dr^2+ r^2 d\Omega_{d-2}^2\,.
\ee

\begin{figure}[h]
\centering
\includegraphics[width=0.6\textwidth]{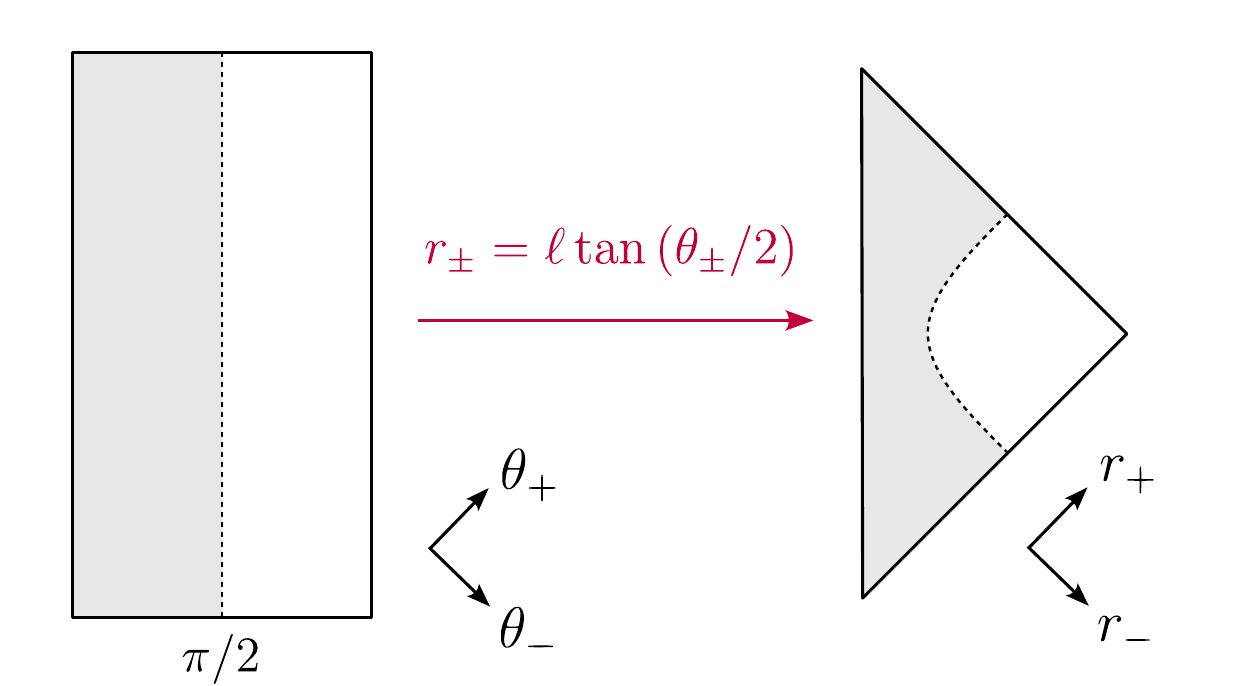}
\caption{Penrose diagrams of the Lorentzian cylinder (left) and Minkowski (right) spacetimes. Both are related via the conformal transformation \eqref{eq:rpmdef}. Moreover, AdS is conformally equivalent to a portion of these spacetimes, as we show in grey.}
\label{fig:conformal_map}
\end{figure}

Comparing (\ref{eq:metricAdS}) and (\ref{eq:dsLC}), we see that AdS$_d$ is conformally equivalent to half of the LC, with a boundary at $\theta=\pi/2$, as shown in the left pannel of Fig. \ref{fig:conformal_map}. Defining a CFT in AdS requires specifying appropriate conformal boundary conditions.\footnote{For a discussion of conformal boundary conditions for free theories see \cite{Aharony:2010ay}.} By the conformal transformation we just reviewed, we can also think about a CFT in AdS as a boundary CFT (BCFT) in the LC or in Minkowski spacetime (see left pannel of Fig. \ref{fig:conformal_map}).\footnote{Using (\ref{eq:rpmdef}), the global AdS boundary at $\theta=\pi/2$ maps to the hyperboloid $r^2-t^2=(2\ell)^2$ in $\mathbb R^{d-1,1}$.} The conformal Killing vectors that keep the boundary fixed give a symmetry group which is the conformal group of a BCFT, namely $SO(d-1,2)$. This is the same as the isometry group of AdS$_d$. Therefore, unlike what happens in the LC or in Minkowski, the conformal group is the same as the isometry group. Then, it is more subtle to distinguish a CFT from a QFT in AdS; this has been recently analyzed in \cite{Carmi:2018qzm}. 

We will consider in AdS the space of QFTs that can be obtained by deforming a CFT by relevant AdS-invariant interactions,
\be
S=S_\textrm{CFT}+ \int d^dx\,\sqrt{-g}\, \lambda_I \mathcal O_I\,,
\ee
where $\mc O_I$ has scaling dimension $\Delta_I <d$. This is the scaling dimension in the BCFT, which can also be measured in AdS by looking at 2-point functions at distances much smaller than the AdS radius $\ell$.

RG flows in AdS can display different dynamics from those in flat space, due to the existence of a dimensionless coupling $\lambda_I^{1/(d-\Delta_I)}\ell$. In the limit of large AdS radius, $\ell \gg \lambda_I^{-1/(d-\Delta_I)}$, the flow is approximately described by the one in flat spacetime: the theory flows between the Minkowski UV and IR fixed points without feeling the curvature, and only at separations of order $\Delta x \sim \ell$ does the AdS geometry modify the correlations. On the other hand, if $\ell < \lambda_I^{-1/(d-\Delta_I)}$, the theory is first affected by the curvature. This effectively creates a gap, since correlators in AdS decay exponentially in terms of the proper distance. The fate of the RG flow in this regime is then model dependent, with some examples studied for instance in \cite{Carmi:2018qzm, Lauria:2023uca, Copetti:2023sya}.

However, it is important to emphasize that this gap is directly related to the exponential growth of the space, and is present even in CFTs. Observables that distinguish a CFT from a QFT should tell the difference between the effects of the AdS scale and those of the RG scale. We will see that quantum information measures can be used to distinguish these cases. In terms of the entanglement entropy, we will provide definitions of ``numbers of degrees of freedom'' that extend those in flat space. These RG charges will be shown to be monotonic under the RG, while they are constant for CFTs in AdS. In other words, the gap created by the AdS geometry will not affect the number of degrees of freedom, but the relevant RG deformations will.

%%%%%%%%%%%%%%%%%%%%%%%%%%%%%%%%%%%%%%%%%
%%%%%%%%%%%%%%%%%%%%%%%%%%%%%%%%%%%%%%%%%
%%%%%%%%%%%%%%%%%%%%%%%%%%%%%%%%%%%%%%%%%
%%%%%%%%%%%%%%%%%%%%%%%%%%%%%%%%%%%%%%%%%
\section{Entanglement structure of CFTs in AdS}\label{sec:markov}

This section is devoted to understanding the entanglement entropy properties of general CFTs in AdS$_d$, which will be crucial in the rest of the manuscript.

The irreversibility formula we will prove in the next section will define irreversible RG charges in terms of the entanglement entropy for the causal diamond of a sphere centered at the origin of AdS. In the global conformal coordinates (\ref{eq:metricAdS}), the entangling region and its associated causal diamond are
\be\label{eq:entangling1}
0\le \theta \le \theta_0\;,\;|T| \le \theta_0 -\theta\,.
\ee
The area of the sphere is determined by
\be
R \equiv \ell\,\tan \theta_0\,.
\ee

We will first establish the Markov property for CFTs in AdS; this result was proved for flat space in \cite{casini2017modular}, and for de Sitter in \cite{Abate:2024xyb, Abate:2024nyh}. This property will then determine the general structure of the entanglement entropy for the vacuum reduced density matrix.

\begin{figure}[h]
\centering
\includegraphics[width=0.3\textwidth]{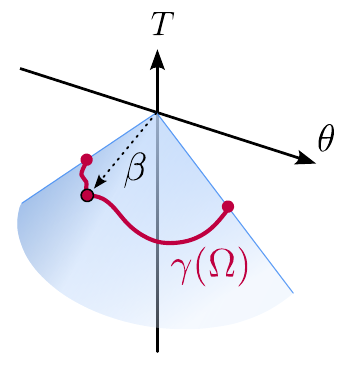}
\caption{Entangling regions with arbitrary boundary $\beta=\gamma(\Omega)$ on the null cone are employed to show the Markov property in AdS. The case $\beta=\beta_0$ gives a spherical region of proper area proportional to $(\ell\tan\beta_0)^{d-2}$.}
\label{fig:wiggly}
\end{figure}

To show the Markov property, we need to include more general entangling regions, corresponding to entangling surfaces that all lie on a common light cone. It is convenient to work with the past light-cone from the origin:
\be\label{eq:nullgeo}
\theta=-T = \beta\;,\;0\le \beta \le \infty\,.
\ee
A spherical entangling region has boundary $\beta= \beta_0$. More general entangling regions in this light-cone are describe by more general curves
\be\label{eq:gammadef}
\beta=\gamma(\Omega)\,,
\ee
as we show in Fig. \ref{fig:wiggly}. The induced geometry on the entangling surface corresponds to a sphere with variable radius,
\be
\dd s^2\Big|_{\gamma}= \ell^2 \,\tan^2 \gamma(\Omega)\, \dd\Omega_{d-2}^2\,.
\ee

Consider the analog situation of entangling regions with boundary on the null cone in flat space-time. In spherical coordiantes,
\be
\dd s^2_\textrm{Mink}= -\dd t^2+\dd r^2 + r^2 d\Omega_{d-2}^2\,,
\ee
the past light-cone from the origin is parametrized by $r=-t = \lambda$, with $0 \le \lambda \le \infty$. The modular Hamiltonian for a CFT and an entagling region with boundary $\lambda= \eta(\Omega)$ on the light-cone was obtained in \cite{casini2017modular},
\be\label{eq:ctt}
H^{\textrm{Mink}}_\eta= 2\pi \int d^{d-2}\Omega\, \int_0^{\eta(\Omega)} d\lambda\, \lambda^{d-1} \frac{\eta(\Omega)-\lambda}{\eta(\Omega)}\, T^{\textrm{Mink}}_{\lambda\lambda}\,,
\ee
where
\be
T^{\textrm{Mink}}_{\lambda\lambda}= \frac{dx^\mu}{d\lambda} \frac{dx^\nu}{d\lambda} T^{\textrm{Mink}}_{\mu\nu}
\ee
is the pull-back of the stress-tensor on the light-cone. So the modular Hamiltonian for wiggly boundaries on the light-cone becomes local when expressed on a light-cone Cauchy surface. This locality is equivalent to the Markov property for the modular Hamiltonian,
\be
H^{\textrm{Mink}}_{\eta_1}+H^{\textrm{Mink}}_{\eta_2}=H^{\textrm{Mink}}_{\eta_1 \cup \eta_2}+H^{\textrm{Mink}}_{\eta_1 \cap \eta_2}\,,
\ee
and this is in turn equivalent to the saturation of strong subadditivity for the corresponding entanglement entropies,
\be
S({\eta_1})+S({\eta_2})= S({\eta_1 \cup \eta_2})+S({\eta_1 \cap \eta_2})\, .
\ee
The saturation of the SSA means that the entanglement entropy will be a local functional of the boundary entangling region geometry, up to a nonlocal universal term. This structure was determined in \cite{casini2018all}. We stress that it is necessary that all the regions involved in the strong subadditivity relation have boundaries in a common light-cone; otherwise the saturation is not valid.

We now use the conformal transformation (\ref{eq:rpmdef}) between flat space and the LC, and the Weyl rescaling between the LC and AdS, to obtain the modular Hamiltonian for CFTs in AdS with boundary entangling regions (\ref{eq:gammadef}) on the past light-cone. 

Let's consider first the effect of the conformal transformation. Recall that a conformal transformation satisfies
\be\label{eq:conf}
\frac{\partial \t x^\mu}{\partial x^\nu}=\Omega(x) \Lambda^\mu_\nu(x)
\ee
where $\Lambda$ leaves the metric invariant, $\Lambda^T g \Lambda = g$. A primary operator of dimension $\Delta$ transforms as
\be
\t \phi_I( \t x) = \Omega(x)^{-\Delta} R_I^{\;J} \phi_J(x)\,,
\ee
where $R$ encodes the spin rotation matrix. For fields in products of the vector representation, the rotation matrix is given by $\Lambda^\mu_\nu$; if we write the transformation law in terms of (\ref{eq:conf}), then a symmetric rank $s$ tensor field transforms as
\be
\t V_{\mu_1 \ldots \mu_s}(\t x)= \Omega(x)^{-\Delta +s} \frac{\partial x^{\nu_1}}{\partial \t x^{\mu_1}} \ldots \frac{\partial  x^{\nu_s}}{\partial \t x^{\mu_s}} V_{\nu_1 \ldots \nu_s}( x)\,.
\ee
For the stress tensor, ignoring anomalies which won't change the operator piece of the modular Hamiltonian, the above formula gives
\be
\t T_{\mu_1 \mu_2}(\t x) = \Omega^{2-d}(x) \frac{\partial x^{\nu_1}}{\partial \t x^{\mu_1}}\frac{\partial x^{\nu_2}}{\partial \t x^{\mu_2}} T_{\nu_1 \nu_2}(x)\,,
\ee
where we used $\Delta=d,\, s=2$. In particular, a null component, defined as
\be
T_{\lambda \lambda} = \frac{dx^\mu}{d\lambda}\frac{dx^\nu}{d\lambda} T_{\mu\nu}
\ee
satisfies
\be\label{eq:Tnull}
\t T_{\lambda \lambda}(\t x) = \Omega^{2-d}(x)T_{\lambda \lambda}( x)\,,
\ee
where the same parameter $\lambda$ is used on both coordinate systems.

On the other hand, a Weyl transformation is a local metric rescaling of the form
\be
\t g_{\mu\nu} = \Omega_W(x)^2 g_{\mu\nu}\,,
\ee
while the coordinates are not changed. From the definition of the stress tensor, $\int \sqrt{g} \;T^{\mu\nu} \delta g_{\mu\nu}= \int \sqrt{\t g}\; \t T^{\mu\nu} \delta \t g_{\mu\nu}$, we obtain
\be
\t T_{\mu\nu}(x)= \Omega_W^{2-d} T_{\mu\nu}(x)\,.
\ee
Under a Weyl transformation, null components are also related as in (\ref{eq:Tnull}).

In summary, the effect of a combined conformal and Weyl transformation on $T_{\lambda \lambda}$ is to rescale it by the overall Weyl factor with a power $2-d$. From (\ref{eq:LCMink}), (\ref{eq:metricAdS}) and (\ref{eq:dsLC}), we obtain
\be
\dd s^2_\textrm{Mink}= \frac{1}{4 \cos^2 \left(\frac{\theta_+}{2}\right) \cos^2 \left(\frac{\theta_-}{2}\right)}  \cos^2 \left(\frac{\theta_+ + \theta_-}{2}\right)\,\dd s_{AdS}^2\,.
\ee
On the past light-cone, $\theta_+=0$, and the Weyl factor is just $1/4$. Therefore
\be\label{eq:Tminktr}
T^\textrm{Mink}_{\lambda \lambda}=2^{d-2}\,T^\textrm{AdS}_{\lambda \lambda}\,.
\ee
From the change of variables $r_\pm= \ell \tan(\theta_\pm/2)$, the Minkowski and AdS null parameters are related by
\be
\tan \beta= \frac{2\lambda}{\ell}\,.
\ee
Performing this change of variables in the modular Hamiltonian (\ref{eq:ctt}) and using (\ref{eq:Tminktr}), we arrive at the modular Hamiltonian for CFTs in AdS with entangling regions on the past light-cone:
\be\label{eq:modHamAdS1}
H^{\textrm{AdS}}_\gamma= 2\pi \ell^{d-2} \int d^{d-2}\Omega\, \int_0^{\gamma(\Omega)} d\beta\, \cos^2\beta\,\tan^{d-1} \beta\, \frac{\tan \gamma(\Omega)-\tan \beta}{\tan \gamma(\Omega)} T^{\textrm{AdS}}_{\beta \beta}\,.
\ee
Here
\be
T^{\textrm{AdS}}_{\beta \beta} = \frac{dx^\mu}{d\beta}\frac{dx^\nu}{d\beta}T^{\textrm{AdS}}_{\mu \nu}=T^{\textrm{AdS}}_{TT}+T^{\textrm{AdS}}_{\theta \theta}\,,
\ee
and we recall that we have subtracted a possible constant (anomaly) piece from the stress tensor.

A quick calculation shows that $\beta$ is not an affine parameter for the null geodesic (\ref{eq:nullgeo}). The affine parameter is given by
\be
\zeta= \tan \beta\,,
\ee
which is proportional to the Minkowski parameter $\lambda$. In terms of this variable, the modular Hamiltonian simplifies to
\be
H^{\textrm{AdS}}_\gamma= 2\pi \ell^{d-2} \int d^{d-2}\Omega\, \int_0^{\tan \gamma(\Omega)} d\zeta\, \zeta^{d-1}\, \frac{\tan \gamma(\Omega)-\zeta}{\tan \gamma(\Omega)}\, T^{\textrm{AdS}}_{\zeta \zeta}\,.
\ee
This result has been obtained before in \cite{Rosso:2019txh}.

The local expression (\ref{eq:modHamAdS1}) implies the Markov property for CFTs with boundary entangling regions along a common past light-cone,
\be\label{eq:markovHAdS}
H^{\textrm{AdS}}_{\gamma_1}+H^{\textrm{AdS}}_{\gamma_2}=H^{\textrm{AdS}}_{\gamma_1 \cup \gamma_2}+H^{\textrm{AdS}}_{\gamma_1 \cap \gamma_2}\,,
\ee
and the saturation of strong subadditivity for the corresponding entanglement entropies,
\be\label{eq:markovSAdS}
S(\gamma_1)+S(\gamma_2)= S(\gamma_1 \cup \gamma_2)+S(\gamma_1 \cap \gamma_2)\,.
\ee

We can now extend the arguments of \cite{casini2018all} to write down the general form of a Markovian EE in AdS. Let $\epsilon$ be a short distance AdS invariant cut-off, and define the induced dimensionless metric on the light-cone boundary,
\be
\hat g_{ab} = \frac{\ell \tan \gamma(\Omega)}{\epsilon}\,g_{ab}\,,
\ee
with $g_{ab}$ the metric for a sphere $S^{d-2}$ with unit radius. Eq. (\ref{eq:markovSAdS}) implies that the EE should be a local functional of the intrinsic boundary geometry plus possibly a nonlocal but Markovian term.\footnote{In principle, one could also include contributions from extrinsic curvatures, but it turns out that these can also be constructed from intrinsic quantities. In particular, the invariant associated with the type B anomaly in $d=4$, which is built using the extrinsic curvatures, vanishes in our setting. The computation that shows these statements parallels the one performed for de Sitter in \cite{Abate:2024nyh}.} Then the most general expression for a markovian CFT EE in AdS is
\be
\label{eq:Sepsilon}
S(\gamma, \epsilon)= \int d^{d-2} \Omega\,\sqrt{\hat g} \left(c_0+c_1 \hat R+c_2 \hat R^2+c_3 (\hat R_{ab}^2+\ldots)\right)+ S_\textrm{non-loc}\,.
\ee
As discussed in \cite{casini2018all}, in odd dimensions $S_\textrm{non-loc}$ is a constant term, which coincides with the universal F-term in the EE. In even dimensions, one can write a bilocal anomaly-type action; its first derivative with respect to the boundary shape is local, and hence the second derivative vanishes and this implies (\ref{eq:markovSAdS}). In particular, for a constant radius entangling surface, this gives
\be\label{eq:nonloc}
S_\textrm{non-loc}= (-1)^{d/2-1} 4 A \log\left( \frac{\ell \tan \gamma}{\epsilon}\right)\,,
\ee
in terms of the A-anomaly.

As an example, the EE for a spherical region in $AdS_4$, with $R= \ell \tan \theta_0$ reads
\be\label{eq:EE4dm0}
S(R)= \mu_1 \frac{R^2}{\epsilon^2}-4 A \log\left( \frac{R}{\epsilon}\right)\,.
\ee
This is formally the same expression as in flat space. In both cases, $R^2$ is the area of the spherical region. But a key difference is that while in flat space $R$ is also the radius of the sphere, in AdS the radius of the sphere $\rho_0$ (proper distance from the center to the sphere) is
\be
R= \ell \sinh(\rho_0)\,.
\ee
For spheres much smaller than the AdS scale, $R \approx \ell \rho_0$ and the EE in AdS behaves as that in flat space in terms of the radius, as expected. However, for $\rho_0 > \ell$, the EE depends exponentially on the radius, since $R \sim \ell e^{\rho_0}$.

This Markov property for CFTs in AdS is one of our key results, and we will next consider more general QFTs in AdS with nontrivial relevant interactions.
In Sections \ref{sec:fermion} and \ref{sec:latticed} we will compute the entanglement entropy for free fermions and bosons using lattice methods. When taking the continuum limit, we will check that the divergence structure coincides with \eqref{eq:Sepsilon} for conformal couplings. When this happens, we will say that the regulator method (a lattice in this case) provides a Markovian cut-off.

%%%%%%%%%%%%%%%%%%%%%%%%%%%%%%%%%%%%%%%%%
%%%%%%%%%%%%%%%%%%%%%%%%%%%%%%%%%%%%%%%%%
%%%%%%%%%%%%%%%%%%%%%%%%%%%%%%%%%%%%%%%%%
%%%%%%%%%%%%%%%%%%%%%%%%%%%%%%%%%%%%%%%%%
\section{Irreversibility formula in AdS}\label{sec:irrev}

In this section we will present the derivation of the irreversibility formula in AdS for QFTs that undergo nontrivial RG flows. We will heavily use AdS invariance, the strong subadditivity property of the entanglement entropy and the Markov property for CFTs in AdS, which we derived in the previous section.

%%%%%%%%%%%%%%%%%%%%%%%%%%%%%%%%%%%%%%%%%
%%%%%%%%%%%%%%%%%%%%%%%%%%%%%%%%%%%%%%%%%
\subsection{Warm-up: the C-theorem in AdS$_2$}\label{subsec:warmup}

In order to develop intuition in the simplest setup, we will first derive the irreversibility formula in $1+1$ dimensions and the C-theorem. Consider the geometric arrangement of boosted geodesics shown in Fig. \ref{fig:bSSA} below.

The geodesic $A$ is given by
\be
\sin(T-T_0)=\tanh\alpha\sin\theta\,,\quad-\theta_0\leq\theta\leq\theta_*\,,
\ee
where $T_0$ and $\theta_*$ are fixed by imposing that the endpoints lie on the null cone $T=\theta_0-\theta$:
\begin{align}
T_0&=\arcsin\of{\tanh\alpha\sin\theta_0}\,,\\
\tan\theta_*&=\frac{\sin(\theta_0-T_0)}{\tanh\alpha+\cos(\theta_0-T_0)}\,.
\end{align}
Note that for $\alpha\to0$ we have $T_0=0$ and $\theta_*=\theta_0$.

The geodesic $B$ is built symmetrically, and the causal union $A\cup B$ coincides with the development of the interval $(-\theta_0,\theta_0)$ at $T=0$. On the other hand, the intersection $A\cap B$ is the causal diamond associated with a straigth line at $T=\theta_0-\theta_*$.

\begin{figure}[h]
    \centering
    \includegraphics[width=0.2\linewidth]{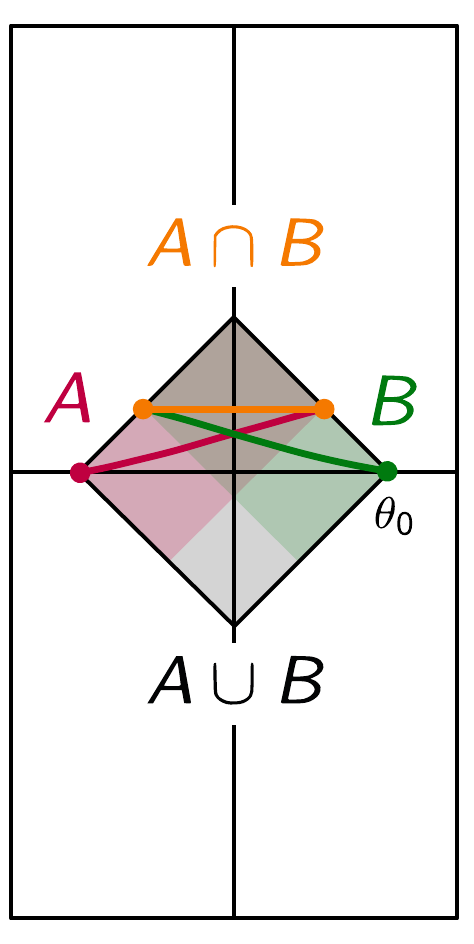}
    \caption{Geometric setup used in the derivation of the infinitesimal boosted SSA inequality.}
    \label{fig:bSSA}
\end{figure}

The proper length of all the geodesics involved can be computed using \eqref{eq:L}:
\begin{align}
    L_A=L_B&=\ell\,\mathrm{arccosh}\of{1+2\tan\theta_0\tan\theta_*}\,,\\
    L_{A\cup B}&=\ell\,\mathrm{arccosh}\of{1+2\tan^2\theta_0}\,,\\
    L_{A\cap B}&=\ell\,\mathrm{arccosh}\of{1+2\tan^2\theta_*}\,.
\end{align}
Note that for $\theta_0\gg1$ we have $L\gg\ell$, diverging in the limit $\theta_0\to\pi/2$.

Now let us plug this into the SSA inequality. This gives a boosted SSA inequality:
\be
S_A+S_B\geq S_{A\cup B}+S_{A\cap B}\ \Rightarrow\ 2S(L_A)-S(L_{A\cup B})-S(L_{A\cap B})\geq0\,.
\ee
In the limit $\alpha\to0$, we obtain an infinitesimal inequality
\be
\partial_{\theta_0}S+\frac{\tan 2\theta_0}{2}\,\partial_{\theta_0}^2 S\leq0\,,
\ee
which can be equivalently written in terms of the geodesic length $L=L_{A\cup B}$ as
\be\label{eq:boostedSSA_L}
\sinh L\,S''(L)+S'(L)\leq0\,.
\ee
In terms of
\be
R= \ell \tan \theta_0 = \ell \sinh(L/2)
\ee
it becomes
\be\label{eq:C-thrm}
R S''(R)+S'(R) \le 0\,.
\ee

This expression is saturated for
\be
S_{\text{Markov}}(L)=c\log\of{R}\,.
\ee
Comparing with (\ref{eq:nonloc}), we see that this corresponds to the EE of a 2d CFT in AdS, where $c$ is proportional to the central charge.  Eq. \eqref{eq:boostedSSA_L}  implies the monotonicity of the following C-function
\be\label{eq:runningC}
C(L)=2\tanh\frac{L}{2}\,S'(L)= R S'(R)\, .
\ee
This then defines a meassure of the relevant number of degrees of freedom at different length scales. For CFTs $C(L)$ is a constant, but for QFTs with RG flows it is not; this RG charge then distinguishes both phases. Furthermore, we see from (\ref{eq:runningC}) that C decreases monotonically as we go to longer distances.
This establishes the C-theorem in AdS, implying the irreversibility of RG flows in AdS$_2$. See also \cite{Meineri:2023mps} for a correlator-based sum rule that implies the C-theorem in AdS$_2$.

%%%%%%%%%%%%%%%%%%%%%%%%%%%%%%%%%%%%%%%%%
%%%%%%%%%%%%%%%%%%%%%%%%%%%%%%%%%%%%%%%%%
\subsection{Proof of the irreversibility formula}\label{subsec:proff}

We will now establish the irreversibility formula for AdS$_d$. This uses the kernel method developed in flat space in \cite{Casini:2023kyj} and also recently used in de Sitter \cite{Abate:2024nyh}. 

As in Sec. \ref{sec:markov}, we work with the past light-cone through the origin,
\be
T=-\theta\,,
\ee
which in the embedding coordinates corresponds to
\be
X^0= \ell\;,\;X^d=- \ell \,\tan \theta\;,\;X^i= \ell \,\tan \theta\, \hat x^i\,.
\ee
We want to consider the effect of isometries that move points in the light cone to points in the light cone. These should leave $X_0$ fixed so we must consider the $J_{MN}$ of $SO(d-1,2)$ with no legs in the $0$ component.

Hence we boost the sphere
\be\label{eq:sphere1}
T= -\theta= - \theta_0\,,
\ee
with a boost transformation $J^{di} \hat n_i$ of boost angle $\eta$%. This keeps $X^0$ fixed, and transforms
\bea
X^d(\eta) &=& \cosh \eta\, X^d + \sinh \eta\,X_n \nonumber\\
X_n(\eta)&=&\cosh \eta\, X_n + \sinh \eta\,X^d\,,
\eea
where we have defined $X_n= \hat n_i X^i$.
Applying the boost to a point on the sphere (\ref{eq:sphere1}) gives
\bea
X^d(\eta) &=& \ell \left(-\cosh \eta\, \tan \theta_0 + \sinh \eta\,\tan \theta_0\,\hat n \cdot \hat x \right)\nonumber\\
X_n(\eta)&=&\ell \left(\cosh \eta\, \tan \theta_0 \,\hat n \cdot \hat x- \sinh \eta\,\tan \theta_0\right)\,.
\eea
Therefore,
\be\label{eq:sphere2}
X^d(\eta)=- \ell \frac{\tan \theta_0}{\cosh \eta}+ \tanh \eta\,X_n(\eta)\,.
\ee
Writing the equation for the boosted sphere as
\be
X^d(\eta)=- \ell \,\tan \theta_\eta(\hat x)\;,\;X^i(\eta)= \ell \,\tan \theta_\eta(\hat x)
\ee
and plugging into (\ref{eq:sphere2}), we obtain
\be\label{eq:boosted_sphere}
\tan \,\theta_\eta(\hat x)= \frac{\tan \theta_0}{\cosh \eta+ \sinh \eta\,\hat n \cdot \hat x}\,.
\ee
This is the same result as in dS$_d$ \cite{Abate:2024nyh}, which we now see then is also valid at negative curvature.

Now we consider $S(\theta_\eta)$, the EE associated with the region with entangling surface determined by the curve $T= -\theta=-\theta_\eta$. Since the vacuum is invariant under boosts we have $S(\theta_\eta)=S(\theta_0)$. Equivalently, under infinitesimal variations
\be\label{eq:constraints}
0=\int\dd\Omega\ \frac{\delta S}{\delta\theta_\eta(\hat x)}\bigg|_{\theta_0}\delta\theta_\eta(\hat x)+\frac{1}{2}\int\dd\Omega_1\dd\Omega_2\ \frac{\delta^2 S}{\delta\theta_\eta(\hat x_1)\delta\theta_\eta(\hat x_2)}\bigg|_{\theta_0} \delta\theta_\eta(\hat x_1)\delta\theta_\eta(\hat x_2)+\dots,
\ee
where the functional derivatives of the EE are associated with derivatives with respect to the proper radius of the spatial sphere $R=\ell\tan\theta_0$ as
\be
\begin{aligned} 
\ell S'(R)&=\cos^2\theta_0\int\dd\Omega\ \frac{\delta S}{\delta\theta_\eta(\hat x)}\bigg|_{\theta_0}\,,\\
\ell^2S''(R)+2\ell\cos\theta_0\sin\theta_0 S'(R)&=\cos^4\theta_0\int\dd\Omega_1\dd\Omega_2\ \frac{\delta^2 S}{\delta\theta_\eta(\hat x_1)\delta\theta_\eta(\hat x_2)}\bigg|_{\theta_0}\,.
\end{aligned}
\ee
Note that rotational invariance imposes that $S_1\equiv\delta S/\delta \theta_\eta|_{\theta_0}$ is a constant, and therefore we have $\ell S'(R)=\cos^2\theta_0 S_1$. It also implies that $\delta^2S/\delta\theta_\eta(\hat x_1)\delta\theta_\eta(\hat x_2)|_{\theta_0}\equiv S_{12}(\hat x_1\cdot x_2)$. 

Plugging \eqref{eq:boosted_sphere} into this expansion we get, order by order in $\eta$, an infinite set of equalities involving integrals of functional derivatives of the EE. To first order we get
\be
S'(R)\int\dd\Omega \of{\hat n\cdot\hat x}=0\,,
\ee
which is automatically satisfied due to rotational invariance. To second order in $\eta$, after some standard manipulations, we get
\be\label{eq:2nd_order} 
\cos\theta_0
\sin\theta_0\int\dd\Omega_1\dd\Omega_2 \of{\hat x_1\cdot\hat x_2}S_{12}(\hat x_1\cdot\hat x_2)-\of{\frac{d-1}{\cos^2\theta_0}-2}\ell S'(R)=0\,.
\ee
Note that, although $S_{12}\leq0$ for non-coincident points due to strong subadditivity, the sign of the first term is not fixed since the integral receives contributions form $\hat x_1=\hat x_2$. 

To proceed, we build the quantity
\be 
\Delta S=S_{\mathrm{QFT}}-S_{\mathrm{UV}}\,,
\ee
where $S_{\mathrm{UV}}$ is the EE for the UV fixed point.\footnote{In $S_{\mathrm{UV}}$ we are also including possible contributions coming from counterterms needed to render the theory free of UV divergences coming from the relevant deformations.} By the Markovian property established in Sec. \ref{sec:markov}, the second order variation $[S_{\mathrm{UV}}]_{12}$ vanishes when the two points do not coincide. On the other hand, the second order variation for $\Delta S$ now vanishes at coincident points, since the entropies for the QFT and the UV fixed point have by construction the same UV behavior. Therefore,
\be
\Delta S_{12} \le 0
\ee
always. Hence the integral
\be\label{eq:int_DeltaS} 
\int\dd\Omega_1\dd\Omega_2\of{1-\hat x_1\cdot\hat x_2}\Delta S_{12}\of{\hat x_1\cdot\hat x_2} \le 0\,.
\ee
Using this inequality into \eqref{eq:2nd_order} in terms of the entropy difference, we arrive to our main result
\be
\label{eq:irrev_formula}
R^2 \,\Delta S''(R)-(d-3) R \,\Delta S'(R) \le 0\,.
\ee

This is the irreversibility formula for AdS. It has the same form as that of dS and $\mathbb{R}^{1,d-1}$ when expressed in terms of the area of the spherical entangling region. However, we must keep in mind that here $R$ grows exponentially with the proper radius of the sphere, unlike dS or $\mathbb{R}^{1,d-1}$. Recalling that $R=\sinh\rho$, we can write \eqref{eq:irrev_formula} as
\be 
\ell\tanh\rho\,\partial_\rho^2\Delta S(\rho)-\of{d-3+\tanh^2\rho}\partial_\rho\Delta S(\rho)\leq0\,,
\ee
where we can explicitly appreciate the non-trivial geometry of AdS. For $\rho\ll\ell$, the last expression reduces to its flat-space counterpart, as expected.

%%%%%%%%%%%%%%%%%%%%%%%%%%%%%%%%%%%%%%%%%
%%%%%%%%%%%%%%%%%%%%%%%%%%%%%%%%%%%%%%%%%
\subsection{RG charges and C, F and A theorems}\label{subsec:theorems}

Let us explore the implications of the irreversibility formula \eqref{eq:irrev_formula} for each $d$ more explicitly. In $d=2$ this gives \eqref{eq:C-thrm}, 
\be
\Delta C(R) \equiv R\,\Delta S'(R)\;,\;\Delta C'(R) \le 0
\ee
establishing the $C$-theorem in AdS. This agrees with the result of Sec. \ref{subsec:warmup}. 

For $d=3$, Eq. (\ref{eq:irrev_formula}) it implies that the quantity 
\be\label{eq:F}
\Delta F(R) \equiv R\,\Delta S'(R)-\Delta S(R)\,,
\ee
is monotonically decreasing along RG flows,
\be\label{eq:monF}
\Delta F'(R) \le 0\,.
\ee
In the UV ($R \to 0$), $\Delta F \to 0$ and $F \to F_{\mathrm{UV}}$, which coincides with the F-charge of the conformal fixed point. We will call this the F-theorem in AdS, establishing the irreversibility of the RG in AdS$_3$.\footnote{To avoid confusions, we should clarify that the recent work \cite{Bason:2025sxb} demonstrated a monotonicity of ``boundary RG flows'' in AdS$_3$, and called this the F-theorem in AdS. In their setup, the bulk QFT does not undergo a nontrivial flow. So this is different from our result (\ref{eq:F}), (\ref{eq:monF}).}

For $d=4$, we can use the second order inequality to define the quantity
\be
\Delta A(R) \equiv \frac{1}{2}\left[ R^2 \,\Delta S''(R)- R \,\Delta S'(R) \right]\,,
\ee
which satisfies
\be\label{eq:DeltaAineq}
\Delta A(R) \le 0
\ee
along the flow. In the UV, the combination
\be
\lim_{R \to 0}\frac{1}{2}\left[ R^2 \, S''(R)- R \, S'(R) \right]=A_{\mathrm{UV}}
\ee
is the A-anomaly of the UV CFT, and $\Delta A \to 0$. Eq. (\ref{eq:DeltaAineq}) is the A-theorem in AdS, implying the irreversibility of the RG in AdS$_4$.

We see that the second order differential inequality allows to define entropic quantities $\Delta C(R),\,\Delta F(R),\,\Delta A(R)$ that are finite, and which measure the number of degrees of freedom relevant at a scale R, relative to the UV fixed point RG-charges $C_{\mathrm{UV}},\,F_{\mathrm UV},\,A_{\mathrm{UV}}$. Operationally, the same definitions apply in flat space. However, there are some key differences. In flat space, the QFT approaches a fixed point in the infrared, and this is detected by $\lim_{R \to \infty} \Delta C=C_{\mathrm{IR}}- C_{\mathrm{UV}}$ (and similarly for $F$ and $A$), which measures the difference in RG charges. This difference is independent of the RG flow. However, in AdS the infrared dynamics need not in general be controlled by the IR CFT that was obtained in the flat space RG, due to effects from the curvature and the asymptotic boundary. This will be detected by our entropic quantities: $\lim_{R \to \infty} \Delta C$ can become a function of the dimensionless combination $M \ell$, where $M$ is a typical energy scale associated to the RG, and $\ell$ is the AdS radius of curvature. In this way, the infrared values of $\Delta C, \Delta F$ and $\Delta A$ could in principle depend on the RG flow. It would be interesting to develop analytic tools to understand the large $R$ limit of these quantum information measures.

Having defined these measures, we can return to the problem of distinguishing conformal and massive theories in AdS, discussed above in Sec. \ref{sec:qftads}. At large distances, both have exponentially suppressed correlators (sometimes called an ``AdS gap''). The RG charges (C, F, A) that we found using entanglement entropy distinguish both phases: they stay constant for a CFT as the entangling radius is increased, while they decrease for a massive QFT.

Finally, we note that for $d >4$, the second order inequality is not powerful enough to cancel the leading $R^{d-k}$ terms in the entropy so as to isolate the universal term. This is the same situation as in flat space. Nevertheless, the inequality still gives nontrivial information. In terms of the area of the entangling region,
\be
a \equiv R^{d-2}
\ee
we obtain
\be
\frac{d^2}{da^2} \Delta S \le 0
\ee
and further (by integrating this),
\be\label{eq:areaterm}
 \frac{d}{da} \Delta S(a_2) - \frac{d}{da} \Delta S(a_1)\le 0\,,\;a_1<a_2\,.
\ee
Eq. (\ref{eq:areaterm}) implies that the coefficient of the area term decreases along the flow. This second inequality is interesting for gravitational questions. The area term change in the EE determines the renormalization of $1/G_N$, when weakly coupling gravity to the QFT. It is also a contribution to black hole entropy from quantum fields. 

%%%%%%%%%%%%%%%%%%%%%%%%%%%%%%%%%%%%%%%%%
%%%%%%%%%%%%%%%%%%%%%%%%%%%%%%%%%%%%%%%%%
%%%%%%%%%%%%%%%%%%%%%%%%%%%%%%%%%%%%%%%%%
%%%%%%%%%%%%%%%%%%%%%%%%%%%%%%%%%%%%%%%%%
\section{Dirac fermion in AdS$_2$}\label{sec:fermion}

In the reminder of this work we will analyze free field theories and their entanglement properties. This will provide explicit examples of the irreversibility formula and the entropic RG charges. In this section we study the entanglement properties of a free massive Dirac fermion in AdS$_2$. The UV fixed point is the conformal theory of a massless Dirac fermion, and the relevant perturbation is the mass term. This will be our first example illustrating the general irreversibility formula we derived before. We will perform both analytic and lattice numerical calculations.

Entanglement entropy for free massive fermions and scalars can be computed exactly in 1+1-dimensional flat space-time; see \cite{Casini:2009sr}. For the Dirac fermion, one approach used bosonization to transform correlators of twist operators of Renyi entropies into sine-Gordon correlators \cite{Casini:2005rm}. These correlations have been computed exactly using form-factor methods, leading to Painlevé equations; see e.g. \cite{Bernard:1994re}. Fortunately for us, these disorder correlation functions have also been computed for Dirac fermions in the hyperbolic plane by \cite{Palmer:1993jp, Doyon:2003nb}. In this section we use these results to express the EE for an interval in AdS$_2$ for a Dirac fermion in terms of the solution to the Painlevé VI equation. We solve this equation numerically and compute the EE and the running C-function, and this provides an explicit check and example of the C-theorem in AdS. We will then compare with lattice calculations, finding excellent agreement. We do not extend the exact calculations to the scalar field, but this would be an interesting line to pursue in the future.

%%%%%%%%%%%%%%%%%%%%%%%%%
%%%%%%%%%%%%%%%%%%%%%%%%%
\subsection{Review of the replica calculation and analytic continuation}
\label{sec:painleve}

For completeness, let us begin with a brief review of the entanglement entropy formula we will use; for more details see \cite{Casini:2005rm, Casini:2009sr}.

For a replicated euclidean manifold $M_n$ branched over a region $V$ (the entangling region), the trace of the n-th power of the reduced density matrix is given in terms of the euclidean partition function over $M_n$ by
\be
\textrm{Tr} \rho_V^n = \frac{Z_n(V)}{Z^n}\,,
\ee
where $Z^n$ is the $n$-th power of the partition function over the unreplicated manifold. The Renyi and von Neumann entropies are
\be
S_n(V) = \frac{1}{1-n} \log \textrm{Tr}\rho_V^n\;,\; S_V = - \textrm{Tr} \rho_V \log \rho_V = \lim_{n \to 1} S_n(V)\,.
\ee

We focus on a fermionic theory with a single Dirac fermion $\psi_j$ in each replica. Combining all the replicated fields into a single $\Psi= (\psi_1, \ldots, \psi_n)$, crossing each cut in the replicated manifold produces a transformation $\Psi \to T_n \Psi$, where
\be
T_n =\begin{pmatrix}
0 & 1 & 0 & \cdots & 0 & 0 \\
0 & 0 & 1 & \cdots & 0 & 0 \\
\vdots & \vdots & \ddots & \ddots & \vdots & \vdots \\
0 & 0 & \cdots & 0 & 1 & 0 \\
0 & 0 & \cdots & 0 & 0 & 1 \\
(-1)^{n+1} & 0 & \cdots & 0 & 0 & 0 
\end{pmatrix}\,.
\ee
This has eigenvalues $e^{2\pi i k/n}$, with $k=- (n-1)/2, \,- (n-1)/2,+1,\,\,\ldots,\,(n-1)/2$, and eigenvectors
\be
\tilde \psi_k = \sum_{j=1}^n e^{2\pi i k/n} \psi_j\,.
\ee

Since the model we are considering is free, the $\t \psi_k$ decouple and the partition function factorizes into a product of partition functions for each of the eigenvectors,
\be
Z_n(V) = \prod_{k=-(n-1)/2}^{(n-1)/2}\,Z_k(V)\,.
\ee
For each of the $\t \psi_k$, going around the cut gives $T_n \t \psi_k = e^{2\pi i k/n} \t \psi_k$ (since they are eigenvectors), and in the Dirac fermion theory this can be obtained by inserting appropriate vortex/antivortex pairs. Dualizing the global $U(1)$ fermionic current as $J_\mu =\epsilon_{\mu\nu} \partial_\nu \phi$, and assuming for concreteness that $V$ is an interval with endpoints $(u, v)$, the partition function is
\be\label{eq:Renyik}
Z_k(V) = \langle e^{-i \sqrt{4\pi} \frac{k}{n} \left( \phi(u)-\phi(v)\right)} \rangle \equiv Z(e^{i 2\pi k/n})\,.
\ee
Then
\be
S_n(V) = \frac{1}{1-n} \sum_{k=-(n-1)/2}^{(n-1)/2}\,\log Z(e^{i 2\pi k/n})\,.
\ee
The entanglement entropy is obtained by analytic continuation $i k/n \to t$ \cite{Casini:2009sr},
\be\label{eq:SV}
S(V) = \pi \int_0^\infty dt\, \frac{\log Z(e^{2\pi t})}{\sinh^2( \pi t)}\,.
\ee

%%%%%%%%%%%%%%%%%%%%%%%%%
%%%%%%%%%%%%%%%%%%%%%%%%%
\subsection{Entanglement entropy in terms of the Painlevé VI solution}

Consider an euclidean Dirac fermion
\be
S= \int d^2x\,\sqrt{g}\, \bar \psi (\gamma^a e_a^\mu \nabla_\mu +m )\psi
\ee
 in the Poincaré disk model for euclidean AdS$_2$ (hyperbolic space) of radius $\ell$. The metric is
\be
ds^2 = \Omega^2 (dx^2+dy^2)\;,\;\Omega=\frac{2\ell}{1-x^2-y^2}\,,
\ee
and so the vielbein reads $e^a_\mu= \Omega \delta_\mu^a$. We choose gamma matrices $\gamma^{a=x}=-\sigma_2$ and $\gamma^{a=y}=\sigma_1$. The spin connection in the covariant derivative
\be
\nabla_\mu= \partial_\mu+ \frac{1}{4} \omega_\mu^{ab}[\gamma_a, \gamma_b]
\ee
becomes
\be
\omega_{\mu=x}^{xy}= \partial_y \log \Omega\;,\;\omega_{\mu=y}^{xy}=- \partial_x \log \Omega\,.
\ee
Rescaling the fermion by $\chi= \Omega^{1/2} \psi$, we arrive at
\be\label{eq:DiracH}
S= \int d^2x\, \bar \chi (\gamma^x \partial_x+\gamma^y \partial_y +\Omega m) \chi\,.
\ee

Ref. \cite{Doyon:2003nb} obtains the 2-point function for the disorder operator $\mathcal O_\alpha$ for the action (\ref{eq:DiracH}), where $\mathcal O_\alpha$ produces a monodromy $\psi \to e^{2\pi i \alpha} \psi$; see also \cite{Palmer:1993jp}. This is the same as a twist operator in the Renyi entropy calculation (\ref{eq:Renyik}) with $\alpha=k/n$. The main result is that this 2-point function is given in terms of the tau function for a Painlevé VI differential system:
\be\label{eq:twistO}
Z_V(e^{i2\pi k/n}) = \langle \mc O_{\alpha=\frac{k}{n}}(u) \mc O_{\alpha'=-\frac{k}{n}}(v) \rangle = \tau(s)
\ee
where
\be\label{eq:sL}
s= \tanh^2\left(\frac{L(u,v)}{2 \ell} \right)\,,
\ee
and $L(u,v)$ is the geodesic distance between the two insertions of disorder operators.

In more detail, Eq. (1.1) of \cite{Doyon:2003nb} gives, after identifying $\alpha=-\alpha'=k/n$ and analytically continuing $i k/n \to t$,
\begin{align}
& w''(s)
-\frac{1}{2}\!\left(
   \frac{1}{w(s)} 
 + \frac{1}{w(s)-1}
 + \frac{1}{w(s)-s}
 \right) \left(w'(s)\right)^2
+ \left(
    \frac{1}{s}
  + \frac{1}{s-1}
  + \frac{1}{w(s)-s}
  \right) w'(s)
\nonumber\\
&\qquad
-\frac{ w(s)\,(w(s)-1)\,(w(s)-s)}{s^{2}(1-s)^{2}}
\left[
    \frac{(1-4\mu^{2})\, s (s-1)}{2 (w(s)-s)^{2}}
  - \frac{s}{2\, w(s)^{2}}
  - 2\, t^{2}
\right]
=0
\label{eq:PV6}
\end{align}
and the tau function satisfies 
\begin{align}
\frac{d}{ds} \log \tau(s) &=\frac{s(1-s)}{4\, w(s)\, (1-w(s))\, (w(s)-s)}
\left[
    w'(s) - \frac{1 - w(s)}{1 - s}
\right]^{2}
- \frac{\mu^{2}}{w(s)-s}
\nonumber\\
&\qquad
- \frac{t^{2}\, w(s)}{s(1-s)}
+ \frac{t^{2}}{s}
+ \frac{\mu^{2} + t^{2}}{1-s}.
\end{align}
In these expressions, $\mu= \ell m$. %The boundary condition for (\ref{eq:PV6}) can be obtained from the long distance limit $s \to 1$.
The boundary condition for $w$ is determined by the properties of a disorder correlator \eqref{eq:twistO}. For short distances one expects a UV conformal behavior while for long distances one imposes cluster decomposition. See Section 7 of \cite{Doyon:2003nb} for details.

Recalling (\ref{eq:SV}), the procedure to compute the entanglement entropy involves first solving  the differential equation (\ref{eq:PV6}) for a given mass $\mu$ and a fixed $t$. The result gives $w(s)$ and hence $\frac{d}{ds} \log Z$ for a given separation $s$, which is related to the proper distance between the interval endpoints by (\ref{eq:sL}).\footnote{For the running C-function, which is our goal here, it is sufficient to have $\frac{d}{ds} \log Z$ and we don't need to perform the $s$-integration.} Then we iterate this procedure for different values of $t$ and perform the $t$-integral in (\ref{eq:SV}) using an interpolating function for the integrand.

\begin{figure}[h]
	\centering
	\includegraphics[width=0.6\textwidth]{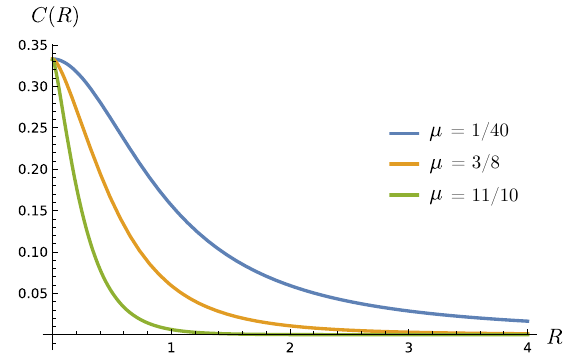}
	\caption{Running C-function for massive Dirac fermion in the hyperbolic disk, evaluated using the Painlevé VI equation.}
	\label{fig:painleve}
\end{figure}

The results of this procedure for the running C-function $C(R)= R S'(R)$ for different masses $\mu=m\ell$ are shown in Fig. \ref{fig:painleve}. This provides explicit examples verifying the C-theorem in AdS.

As discussed in \cite{Casini:2005rm, Casini:2009sr}, the Renyi entropies for scalar fields in flat space-time can also be obtained from sine-Gordon correlators; this has been established in \cite{Delfino:2002dd}. It would be an interesting future direction to extend these methods to AdS$_2$ as well.

%%%%%%%%%%%%%%%%%%%%%%%%%
%%%%%%%%%%%%%%%%%%%%%%%%%
\subsection{Lattice calculations}\label{subsec:Diraclattice}

We will now compute the entanglement entropy and running C-function numerically, by placing the Dirac fermion in an AdS$_2$ lattice. The lattice is obtained by discretizing the spatial direction. 

Lattice calculations in AdS bring in new challenges compared to the flat space case. One of them is to ensure that the AdS isometries emerge in the continuum limit. Another important aspect is to check the Markov property holds in the continuum limit for the conformal fixed point. Explicit calculations show that the choice of spatial coordinates and its associated lattice cut-off can have a big impact on such issues. 

For the Dirac fermion in AdS$_2$ we have the advantage of already knowing the continuum results for the entanglement entropy in terms of the Painlevé solution. As we illustrate below, the best agreement with the continuum answer is obtained for the global AdS$_2$ metric with proper distance coordinate,
\be\label{eq:metricAdS2rho}
\dd s^2=\ell^2\of{-\cosh^2 \rho\,\dd T^2+\dd \rho^2}\,,
\ee
where $\rho \in (-\infty, \infty)$. So in what follows we focus mostly on this coordinate choice.

We discretize the spatial direction,
\be
\rho_j= \epsilon_\rho\,j\,.
\ee
A ``good'' lattice must regulate gradient fluctuations at distance scales much smaller than the RG scale (in our case, the inverse mass): $\epsilon_\rho^2 \ll 1/(\ell^2 m^2)$. It should also recover the AdS isometries in the continuum limit. This makes $\rho$ a natural coordinate choice, since one of the AdS `translations' acts linearly on it, $J_{01}\big|_{T=0}=\partial_\rho$. As in flat space, a way to test this emergence of AdS invariance is to evaluate scalar lattice quantities and check that in the continuum limit they depend only on the geodesic length.

We start from the action for a Dirac fermion in AdS and then discretize it. We follow the conventions for curved space-time fermions of \cite{Parker:2009uva}. Here it is better to work in signature $(+-)$, where the action reads
\be\label{eq:Ssym1}
S= \int d^2x \sqrt{-g}\left[\frac{i}{2} (\bar \psi \gamma^a e_a^\mu D_\mu \psi- D_\mu \bar \psi \gamma^a e_a^\mu \psi)- m \bar \psi \psi\right]\,.
\ee
We choose the flat space gamma matrices to be
\be
\lbrace \gamma^a, \gamma^b \rbrace = 2 \eta^{ab}\;,\;\gamma^0 = \sigma^1\,,\;\gamma^1= i \sigma^2\,,
\ee
$e^a_\mu$ is the vielbein and the covariant derivative is given in terms of the spin connection by
\be
D_\mu \psi= \partial_\mu \psi + \frac{1}{8} \omega_\mu^{ab} [\gamma_a, \gamma_b] \psi\,.
\ee
It is simplest to first work out explicitly the action in global conformal coordinates.
A short calculation gives
\be
S=\ell  \int dT d\theta\, \frac{1}{\cos^2 \theta} \left[i \cos \theta \left(\bar \psi \sigma^1 \partial_0 \psi+ i \bar \psi \sigma^2 \partial_1 \psi \right)- \frac{1}{2} \sin \theta \,\bar \psi \sigma^2 \psi- \ell m \bar \psi \psi \right]\,.
\ee
It is useful to redefine the fermionic field to the canonical field
\be
\psi(T ,\theta)= (\cos \theta)^{1/2}\,\t \chi(T, \theta)\,.
\ee
Furthermore, denoting the two spinor components by $\t \chi_+$ and $\t \chi_-$, the component action becomes
\be
S=\ell  \int dT d\theta\,  \left[i  \t \chi_+^*(\partial_T-\partial_\theta) \t \chi_++i  \t \chi_-^*(\partial_T+\partial_\theta) \t \chi_-- \frac{\ell m}{\cos \theta} ( \t \chi_+^* \t \chi_-+\t \chi_-^* \t \chi_+) \right]\,.
\ee
This exhibits the two chiralities, coupled by a mass term which becomes position-dependent due to the AdS geometry.

Now changing variables to the proper distance coordinate, $\tanh \rho= \sin \theta$, and rescaling the fermions so that they have a canonical kinetic term (call the new fields $\chi_i$), we arrive at
\be
S=\ell \int dT d\rho \left[i \chi_\pm^* \partial_T \chi_\pm \mp \frac{i}{2} \cosh \rho (\chi_\pm^* \partial_\rho \chi_\pm- \partial_\rho \chi_\pm^*\,\chi_\pm)-\ell m \cosh \rho (\chi_+^* \chi_-+\chi_-^* \chi_+) \right]
\ee
We have written the antisymmetrized spatial gradient to simplify the discretization below.

The discretization is done now as in flat space-time, the only difference being that the mass term is position-dependent. For a review of this procedure in our context, see the Appendix of \cite{Casini:2005rm}. Due to the doubling of low energy fermion modes on the lattice, it is sufficient to consider a spinless fermion $\chi$. Latticizing the proper distance as 
\be
\rho_j= \epsilon \left(j-\frac{N+1}{2} \right)\;,\;j=0,\,\ldots,\,N+1\,,
\ee
and imposing Dirichlet boundary conditions\footnote{In terms of two-point functions that approach the boundary of AdS, this corresponds to the standard quantization in the context of AdS/CFT. For small enough masses it is also possible to choose an alternative quantization, but we don't explore this possibility here.}
\be
\chi_{j=0}=\chi_{j=N+1}=0\,,
\ee
the Hamiltonian becomes
\be\label{eq:spinless3}
H= \sum_{j=1}^N\,\cosh\left(\rho_j \right) \left[\frac{i}{2}(\chi_{j+1}^* \chi_j- \chi_j^* \chi_{j+1})-(-1)^j \epsilon \,\ell m\,\chi_j^* \chi_j\right]\,.
\ee
To arrive to this expression, we took discrete derivatives $\partial_\rho \chi \to \frac{1}{\epsilon}(\chi_{j+1}-\chi_j)$; furthermore, the oscillating term proportional to $(-1)^j$ for the spinless fermion gives the mass term for the Dirac fermion.

Let us now present the numerical results for the entanglement entropy for the spinless lattice fermions with the above Hamiltonians. We follow the numerical method reviewed in detail in \cite{Casini:2009sr}, which computes the eigenvalues of the reduced density matrix (and from there the EE) in terms of the eigenvalues and eigenvectors of equal time 2-point functions. After evaluating the entanglement entropy, we compute the running C- function $C(R) = R S'(R)$ defined in Eq. (\ref{eq:runningC}). This has the advantage of cancelling the leading cut-off dependence from the entanglement entropy, which diverges logarithmically in the continuum limit.

We have performed lattice calculations for different values of the cut-off $\epsilon$ and the finite size $N$. For the running C-function, we have to take small enough $\epsilon$ such that  further decreasing $\epsilon$ does not affect $C(R)$; this ensures convergence to the continuum limit. Similarly, different values of $N$ are taken in order to extrapolate to $N \to \infty$. We have considered values up to $\epsilon=10^{-3} $ and $N=6000$. We also compute the entanglement entropy and running C-function for intervals centered at different positions, in order to check that the results only depend on the geodesic length of the interval. 

\begin{figure}[h]
    \centering
    \includegraphics[width=0.6\textwidth]{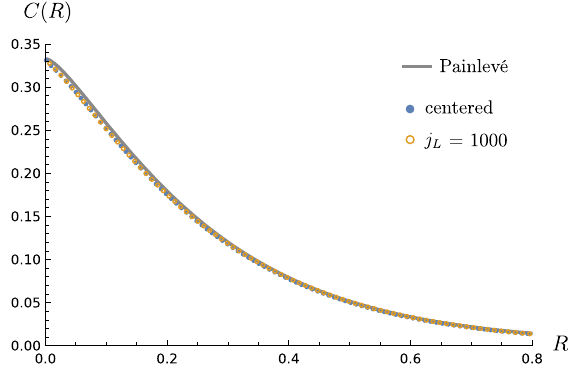}
    \caption{Running C-function for the $\rho$ lattice, for $\ell m=11/10$, $\epsilon=10^{-3}$ and $N=6000$ computed for an interval centered at the origin and another interval with left endpoint fixed at the lattice site $j_L=1000$. We find excellent agreement between the theoretical Painlevé result and the lattice calculations. AdS translation invariance along $\rho$ is obeyed to very good approximation since the centered and non-centered intervals give the same result.}
    \label{fig:CRrho}
\end{figure}

In Fig. \ref{fig:CRrho} we present lattice results and compare them to the analytic (Painlevé) predictions. We obtain an excellent agreement between the theoretical and numerical approaches, and verify that the AdS translation symmetry emerges in the continuum limit. This consistency between the two approaches is quite significant, both as a check of the AdS Painlevé results (which as far as we know were not compared to numerics before), as well as for validating our lattice calculations. This same lattice methodology is extended in the next section to higher dimensions, where so far no analytic results are known.\footnote{Lattice calculations in the global conformal coordinate $\theta$ for similar values of $N$ do not produce such a good agreement, and exhibit noticeable violations of AdS invariance in the continuum limit.}

%%%%%%%%%%%%%%%%%%%%%%%%%%%%%%%%%%%%%%%%%
%%%%%%%%%%%%%%%%%%%%%%%%%%%%%%%%%%%%%%%%%
%%%%%%%%%%%%%%%%%%%%%%%%%%%%%%%%%%%%%%%%%
%%%%%%%%%%%%%%%%%%%%%%%%%%%%%%%%%%%%%%%%%
\section{Lattice calculations for free scalar fields}\label{sec:latticed}

Now we begin a systematic analysis of lattice calculations for the EE of free bosonic theories in AdS$_d$. The renormalization group flow corresponds to starting from a conformally coupled scalar, and adding a relevant mass deformation. All the information is in the scalar 2-point function, which can be computed explicitly; so this yields a very simple and solvable example of an RG flow. Nevertheless, the entanglement properties turn out to be quite nontrivial. First we will explore the lattice version for the theory in arbitrary dimensions, and then we will focus on numerical computations of the C, F and A-functions separately.

\subsection{Scalar in AdS$_d$}

Consider a real scalar field $\phi$ conformally coupled to the curvature. As in the previous section we will work on global coordinates \eqref{eq:global_ads}, where the action takes the form
\be
S=\frac{\ell^{d-2}}{2}\int\dd^d x\, \cosh\rho\,\sinh^{d-2}\rho\of{\frac{(\partial_t\phi)^2}{\cosh^2\rho}-(\partial_\rho\phi)^2+\frac{\phi\nabla_\Omega^2\phi}{\sinh^2\rho}-\ell^2 m^2\phi^2}\,.
\ee
Here $\nabla_\Omega^2$ represents the Laplacian over the $(d-2)$-dimensional unit sphere, and we are including the conformal coupling in the mass term, namely
\be 
m^2=m_0^2-\frac{d(d-2)}{4\ell^2}\,;
\ee
thus $m_0=0$ represents the conformal fixed point. 

Introducing the functions $Y_{l\vec{m}}$, which are normalized eigenfunctions of the $d-2$ sphere Laplacian with eigenvalue 
\be\label{eq:eigenvalues_Ylm}
k_l^2=-l(l+d-3)\,,\ l\in\mathbb{N}_{\geq0}\,,
\ee
we can decompose the field as 
\be\label{eq:modes}
\phi(t,\rho,\Omega)=\of{\frac{\cosh\rho}{\ell^{d-2}\sinh^{d-2}\rho}}^{1/2}\sum_{l,\vec{m}}\chi_{l\vec{m}}(t,\rho)Y_{l\vec{m}}(\Omega)\,,
\ee
so the action for each mode decouples as $S=\sum_{l,\vec{m}}S_{l\vec{m}}$. Moreover, integrating the angular part yields 
\be\label{eq:action_scalar_d}
S_{l\vec{m}}=\frac{1}{2}\int\dd t\,\dd\rho\off{(\partial_t\chi_{l\vec{m}})^2-\mu_d\of{\partial_\rho(f_d\chi_{l\vec{m}})}^2-\ell^2 m_{\mathrm{eff}}^2\chi_{l\vec{m}}^2}\,,
\ee
where we have defined 
\be 
\mu(\rho)=\cosh\rho\,\sinh^{d-2}\rho\ ,\quad f(\rho)=\frac{\cosh^{1/2}\rho}{\sinh^{(d-2)/2}\,\rho}\ ,
\ee
and $m_{\mathrm{eff}}^2$ is an effective, space-dependent coupling with mass units 
given by
\be
m^2_{\mathrm{eff}}(\rho)=m^2\cosh^2\rho- \frac{l(l+d-3)}{\ell^2\,\tanh^2\rho}\,.
\ee
The prefactor in \eqref{eq:modes} is chosen to render the action \eqref{eq:action_scalar_d} canonically normalized. 

In order to have a well-posed variational problem from this action, we also have to provide  boundary conditions for the field at the AdS$_d$ boundary. We will work with Dirichlet conditions at the asymptotic boundary (appropriately regularized). The other boundary condition arises as usual as a regularity condition in the interior. In more detail, the solutions near the origin behave as
\be\label{eq:origin_behaviour}
\chi_{l\vec{m}}\sim\rho^{d/2-1}\of{A\rho^l+B\rho^{3
-d-l}}\,.
\ee
The regular solution has $B=0$, and hence $\chi_{l\vec{m}}(0)=0$.

Now we introduce a lattice spacing $\epsilon$ and discretize the radial variable as
\be
\rho_j=\epsilon j\,,\quad j=1,\dots,N\,.
\ee
The derivative appearing in \eqref{eq:action_scalar_d} is also modified according to
\be 
\partial_\rho\chi_{l\vec{m}}\to\frac{\chi_{l\vec{m}}(j+1)-\chi_{l\vec{m}}(j)}{\epsilon}\,,
\ee
Using this one can write a discretized version of the Hamiltonian as
\be \label{eq:lattice_hamiltonian1} 
H=\frac{\epsilon}{2}\sum_{l,\vec{m}}\sum_j\off{\pi_{l\vec{m}}^2(j)+\mu(j)\of{\frac{f(j+1)\chi_{l\vec{m}}(j+1)-f(j)\chi_{l\vec{m}}(j)}{\epsilon}}^2+\ell^2 m_{\mathrm{eff}}^2(j)\chi_{l\vec{m}}^2(j)}\,,
\ee
for $\pi_{l\vec{m}}=\partial_t\chi_{l\vec{m}}$, by virtue of the canonical normalization. Also, we are using the following prescription for the $\mu$ and $f$ factors:
\be
\mu(j)\equiv\mu(\rho_j+\epsilon/2  )\,,\quad f(j)\equiv f(\rho_j)\,.
\ee
A slightly different lattice model can be obtained by first integrating by parts the cross-terms $\chi \partial_\rho \chi$ in (\ref{eq:action_scalar_d}), and then latticizing $(\partial_\rho \chi)^2$; we have checked that this alternative lattice gives worse numerical convergence than (\ref{eq:lattice_hamiltonian1}).\footnote{In flat space there is a similar ambiguity in the choice of lattice; see e.g. \cite{Lohmayer:2009sq}.}
Furthermore, we note that the factor of $\mu(j)$ is placed at the midpoints between the two sites contributing to the gradient term; this ensures that this term is symmetric. Other symmetric choices are possible, but this one gives faster numerical convergence.

Now we come to the issue of boundary conditions in the lattice model. As we will discuss in more detail in each dimensionality below, AdS entanglement entropies turn out to be more sensitive to boundary conditions than their flat space counterparts. We will mimic the asymptotic boundary condition by adding an extra site at $j=N+1$ and requiring the Dirichlet boundary condition
\be
\chi_{j=N+1}=0\,.
\ee
This means that the sum in (\ref{eq:lattice_hamiltonian1} ) runs up to $j=N$, and we set to zero the terms proportional to $\chi_{N+1}$. This is a finite size approximation, since in the continuum limit the Dirichlet boundary condition is set at $\rho \to \infty$.

Regarding the boundary condition at the origin, there are two natural possibilities: (i) fixing the field to a specific choice, for instance a Dirichlet boundary condition, corresponding to adding a site at $j=0$, setting $\chi_{j=0}=0$ and starting the sum from $j=0$; and (ii) a ``free'' boundary condition, corresponding to simply starting the sum at $j=1$ and letting $\chi_{j=1}$ adjust. From the point of view of the continuum limit, this second choice seems better, since at the origin we have a regularity condition and not a special point where we pin the field to a particular value. We have numerically explored both, and have found that the free boundary condition has a better continuum limit. So we will focus in this one.

To summarize, the lattice model becomes
\be\label{eq:lattice_hamiltonian2} 
H=\frac{\epsilon}{2}\sum_{l,\vec{m}}\sum_{i,j=1}^N\off{\pi_{l\vec{m}}^2(j)+\chi_{l\vec{m}}(i)K^{(l\vec{m})}_{ij}\chi_{l\vec{m}}(j)}\,,
\ee
where the hermitian matrix $K^{(l\vec{m})}$ is obtained from (\ref{eq:lattice_hamiltonian1} ). We will deal with this for each $d=2,3,4$ in more detail in the following subsections. Having obtained the matrix $K^{(l\vec{m})}$, the entanglement entropy for the system is calculated as described e.g. in \cite{Peschel:2002yqj,Casini:2009sr}.  %, which we just review in the following.

\begin{comment}
\textcolor{red}{**Este párrafo tal vez se puede omitir y simplemente poner citas**}
One key observation is that, since each angular mode is decoupled in \eqref{eq:action_scalar_d}, the vacuum state is just a product state. Moreover, the two point functions of the field and momentum between different modes vanishes, so in the following we will drop the indices $l,\vec{m}$. One can obtain these functions, namely $X_{ij}\equiv\langle\chi(i)\chi(j)\rangle$ and $P_{ij}=\langle\pi(i)\pi(j)\rangle$, in terms of $K$ as
\be 
X=\frac{1}{2K^{1/2}}\,,\quad P=\frac{K^{1/2}}{2}\,.
\ee 
Since we are dealing with a gaussian theory, the information of the state is completely encoded in these quantities. More precisely, if we consider a subset $V$ on the lattice and build the reduced matrices $X_V$ and $P_V$. The regions we will consider in this work are of the form $V=\{\rho_1,\dots,\rho_{j_{\max}}\}$ for some $j_{\max}\leq N_{\max}$, representing spheres centered in the origin in the continuum theory. It turns out that the spectra of the state reduced to $V$ is encoded in $C_V=\sqrt{X_VP_V}$, which is an hermitian matrix with eigenvalues in the range $(-1/2,1/2)$. Furthermore, one can compute the EE of the region $V$ for each angular mode as
\be
S_l(V)=-\mathrm{tr}\off{(C_V-1/2)\log\of{C_V-1/2}+(C_V+1/2)\log\of{C_V+1/2}}\,.
\ee
\end{comment}
Since the model is free, the entanglement entropy simply reduces to the sum of entropies for each mode%Finally, the full entropy is obtained as the sum over every mode $l$, weighted by the degeneracy in $\vec{m}$ as 
\be
\label{eq:EEl}
S=\sum_{l=0}^\infty\lambda_l(d)S_l, 
\ee
weighted by the corresponding degeneracy
\be 
\lambda_l(d)=(2l+d-3)\frac{(l+d-4)!}{l!(d-3)!}\,.
\label{degfor}
\ee
In practice, the sum is done up to a cut-off $l_{\max}$ and then it is extrapolated to $l_{\max} \to \infty$.\footnote{It is also possible to fit the $S_l$ for large $l$ and sum these fits up to infinity; see e.g. \cite{Huerta:2022tpq}. In the results below, we checked that both strategies yielded the same results, although the latter usually exhibited more numerical noise.}

In the following subsections we will evaluate numerically \eqref{eq:EEl} in dimensions $d=2,3,4$ and compute the corresponding running RG charges. For the sake of numerics we will set $\ell=1$ and present our results by measuring the mass and $R$ in units of the AdS radius.

\subsection{Entanglement in AdS$_2$}\label{subsec:Ed2}

The case $d=2$ is particular because there are no angular directions and $\rho \in(-\infty, \infty)$. We will consider two approaches and contrast them to detect possible issues associated to the discretization scheme. One option is to discretize the $\rho$ line, introducing two boundaries. Another option is to recognize that due to the parity symmetry $\rho \to -\rho$, the eigenvectors can be dividied into even and odd functions. It is then enough to consider $0\le \rho \le \infty$ and sum over even and odd modes which obey, respectively, Neumann and Dirichlet boundary conditions at the origin.\footnote{
The mode expansion \eqref{eq:modes} can be adapted to this case, noticing that the eigenvalues of the spherical harmonics \eqref{eq:eigenvalues_Ylm} vanish in $d=2$ by taking $l=0$ or $l=1$. So \eqref{eq:modes} implies that the field can be expanded as a combination of two orthogonal fields $\chi^{(l=0)}$ and $\chi^{(l=1)}$, with linear behavior at the origin according to \eqref{eq:origin_behaviour}. Either the solutions themselves or their derivatives vanish at the origin. Thus we can still work on the half-line $\rho>0$ by identifing the modes for $l=0,1$ with configurations attaining Neumann or Dirichlet boundary conditions at $\rho=0$, respectively.}

Let us begin with the computations for the lattice where $\rho$ ranges over a half line. 
As discussed before, we impose a Dirichlet boundary condition at the infrared cutoff, namely $\chi_{j=N+1}=0$. For odd modes (formally $l=1$) in the full line (formally , we impose the Dirichlet boundary condition at the origin, by adding an extra $j=0$ site in the lattice, demanding that
\be
\chi^{(l=1)}_{j=0}=0\,,
\ee
and starting the lattice sum from the $j=0$ term.
Replacing this in \eqref{eq:lattice_hamiltonian1},  we can write the Hamiltonian for the system as \eqref{eq:lattice_hamiltonian2} with
\be\label{eq:K_d=2_dirichlet} 
\begin{aligned}
K^{(l=1)}_{j,j}&=\off{\mu(j)+\mu(j-1)}f^2(j)+\epsilon^2 m_{\mathrm{eff}}^2(j)\,,\\
K^{(l=1)}_{j,j+1}&=K^{(l=1)}_{j+1,j}=-\mu(j)\sqrt{f(j+1)f(j)}\,,
\end{aligned}
\ee
the remaining elements vanishing. Here $m_{\mathrm{eff}}(j)\equiv m_{\mathrm{eff}}(\rho_j)$.

Meanwhile, for the even ($l=0$) modes we impose a Neumann boundary condition by adding an extra site at $j=-1$ and demanding
\be
\chi^{(l=0)}_{j=-1}=\chi^{(l=0)}_{j=1}\,,
\ee
since we expect to obtain an even mode in the continuum. Using this, the Hamiltonian can again be cast as in \eqref{eq:lattice_hamiltonian2} but beginning the sum at $j=0$ and taking
\be
\begin{aligned}
K^{(l=0)}_{0,0}&=\off{\mu(1)+\mu(0)}f^2(0)+\epsilon^2m_{\mathrm{eff}}^2(0)\,,\\
K^{(l=0)}_{1,1}&=\off{2\mu(1)+\mu(0)}f^2(1)+2\epsilon^2m^2_{\mathrm{eff}}
(1)\,,
\end{aligned}
\ee
and \eqref{eq:K_d=2_dirichlet} for $j\geq2$.
%\be
%\begin{aligned}
%K^{(1)}_{0,0}&=\cosh\of{3\epsilon/2}+1+\epsilon^2m^2\,,\\
%K^{(1)}_{1,1}&=\off{2\cosh\of{3\epsilon/2}+1}\cosh\epsilon+2\epsilon^2m^2\cosh^2\epsilon\,,\\
%K^{(1)}_{j,j}&=\off{\cosh\of{\epsilon(j+1/2)}+\cosh\of{\epsilon(j-1/2)}}\cosh\epsilon j+\epsilon^2m^2\cosh^2\epsilon j\,,\\
%K^{(1)}_{j,j+1}&=K^{(1)}_{j+1,j}=-\cosh\of{\epsilon(j+1/2)}\sqrt{\cosh\of{\epsilon (j+1)}\cosh\epsilon j}\,,
%\end{aligned}
%\ee

Regarding the alternative lattice defined where $\rho$ ranges over the whole line, we now discretize as $\rho_j=\epsilon(j-N-1)$ for $j=1,\dots,2N+1$ sites and consider intervals centered at the origin, i.e. at $j=N+1$. The Hamiltonian is still of the form \eqref{eq:lattice_hamiltonian2} and the matix $K_{ij}$ coincides with \eqref{eq:K_d=2_dirichlet} simply dropping the centrifugal term in $m_{\mathrm{eff}}$.

\begin{figure}[h]
    \centering
    \includegraphics[width=0.6\linewidth]{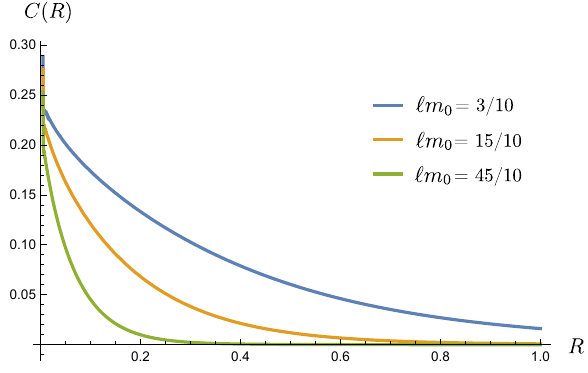}
    \caption{C-function for different values of the mass parameter in units of the AdS radius, $\epsilon=1/500$ and $N=5000$, computed as the sum of the contributions from the $l=0$ and $l=1$ modes.}
    \label{fig:Cbcompare}
\end{figure}

In Fig. \ref{fig:Cbcompare} we show the C-function $C(R)=RS'(R)$ for $m_0=3/10$, $m_0=15/10$ and $m_0=45/10$ in units of the AdS radius, obtained using the mode decomposition. In all cases we used $N=5000$ and $\epsilon=1/500$, but we only display the result up to $j_{\max}=500$ in order to mitigate the finite-size effects. We performed several convergence tests in $N$ and $\epsilon$ aiming at the continuum theory, finding that our results are converged for the values reported. The curves are obtained by interpolating the results for the EE before performing derivatives, avoiding the noise inherent to numerical differentiation. 

Note that all the C-functions are monotonically decreasing, in agreement with our irreversibility formula \eqref{eq:irrev_formula}. Near $R=0$ we can see a cusp attributed to the IR divergence of the zero mode of the massless theory. This is the case for the flat-space C-function \cite{Casini:2009sr}, and we will soon confirm this in AdS$_2$ as well. This cusp seems to approach the expected value of $1/3$, since the central charge of the UV theory is $C_{\mathrm{UV}}=1$, but our current resolution does not allow us to take $R\leq0.02$. In the opposite limit we see that the C-functions approach to zero, as expected for a gapped theory, faster as we increase the mass parameter.

\begin{figure}[h]
    \centering
    \includegraphics[width=0.6\linewidth]{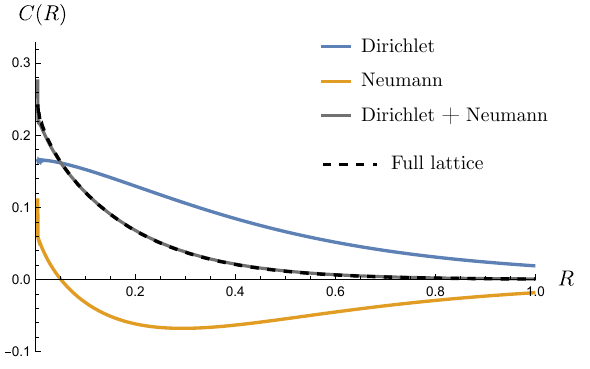}
    \caption{Comparison between the C-functions of the two discretization schemes presented for $\ell m_0=15/10$, as well as the ones obtained for the Dirichlet and Neumann modes. All curves were computed using $\epsilon=1/500$ and $N=5000$.}
    \label{fig:Cb15o10}
\end{figure}

In Fig. \ref{fig:Cb15o10} we compare the C-function for $m_0=15/10$ previously shown, with the one obtained using the lattice over the whole line. Note that both lattices exhibit excellent agreement, which serves as a crosscheck of our discretization method. We also display the contribution for each of the Dirichlet and Neumann modes individually. We observe a cusp near $R=0$ in the C-function for the Neumann mode, which is absent in the Dirichlet case. The cusp is associated with the IR divergence of the massless field zero mode. Meanwhile, the latter smoothly approaches $1/6$ as $R\to0$, which is expected for this theory with boundary \cite{calabrese2004entanglement}. 

Furthermore, we note that the C-function for the Neumann mode is not monotonic. We remark that this is not a contradiction, since we have not established an irreversibility theorem for theories with defects in AdS, but we think that this is an interesting line of research in the future. These configurations conformally map to a Casimir problem, where irreveribility theorems are not yet known.
 
\subsection{Entanglement in AdS$_3$}\label{subsec:Ed3}

In this section, we will analyze the $2+1$ dimensional case making emphasis on the particularities found for this dimension.
To begin with, we also have to introduce an extra cut-off $l_{\max}$ corresponding to the maximum value of $l$ we will consider in the sum \eqref{eq:EEl}. Also, the degeneracy \eqref{degfor} gives $2$ for $l>1$ and $1$ for $l=0$.
This sum over $l$ will effectively produce a area law term. We will check that our numerics for $d\geq 2$ reproduces it, consistently with a Markovian cutoff.

Another particularity is that the universal part of the entropy at fixed points is a constant. This makes numerical calculations of $F$ notoriously hard, because both lattice and boundary condition effects strongly affect the constant term in the entropy. For a discussion of cutoff effects on $F$ see e.g. \cite{Casini:2015woa}. Regarding the boundary condition at the origin, we compared the free boundary condition (corresponding to starting the sum from $j=1$) with for instance a Dirichlet boundary condition. We verified that the free one gives the correct results in the continuum limit; this holds both in AdS and in flat space.\footnote{We thank Marina Huerta for very helpful discussions about numerical calculations of F in flat space.} On the other hand, we found that small changes in the boundary condition at the infrared cutoff $j=N$ do not lead to appreciable modifications in the entanglement entropy behavior.

We work with double precision, which enables the use of fast libraries for linear algebra. These are needed for diagonalizing the quadratic matrix $K$ in (\ref{eq:lattice_hamiltonian2}) and then constructing the reduced density matrix for each interval (using the boson 2-point function). Working in units of the AdS radius $\ell=1$, we have considered lattice cutoffs $\epsilon$ in the range $10^{-1}$ -- $10^{-2}$, lattice sizes $N$ in the range $100$ to $3000$, and angular momentum cutoffs $l_\textrm{lmax}$ up to $30 \,\times\,10^3$. Due to the exponential factors in $K$, we can achieve proper sizes $\rho \sim 10$ given our working precision; this translates into area factors of order $R \sim 10^5$, which is enough for our purpose. Furthermore, we have studied masses $\ell m$ in the range $0$ -- $5$. Similar numerical ranges are also used in AdS$_4$ below.

\begin{figure}[h]
    \centering
    \includegraphics[width=0.6\linewidth]{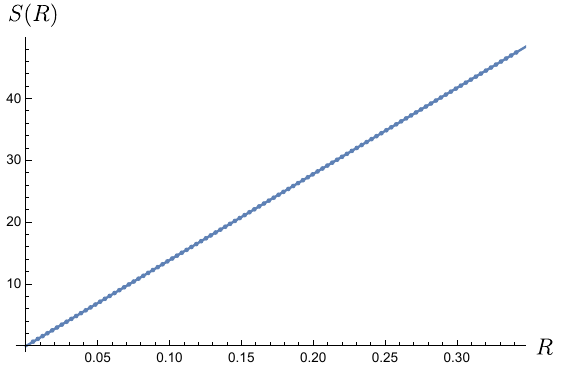}
    \caption{Entanglement entropy as a function of $R$ for the conformally coupled scalar, for $\epsilon=1/300$, $N=3000$, $l_\textrm{max}=15\,\times\,10^3$. Dots are the numerical results, and the line is a global linear fit Eq. (\ref{eq:linear1}). }
    \label{fig:S3d1}
\end{figure}

Let us present first the numerical results for the conformally coupled scalar, which corresponds to the UV fixed point of the massive RG flow. In Fig. \ref{fig:S3d1} we show the numerical results for the entanglement entropy of the conformally coupled scalar; the dots are obtained for a lattice with  $\epsilon=1/300$, $N=3000$, adding up to $l_\textrm{max}=15\,\times\, 10^3$. Within our working precision, these results were found to be converged under decreasing $\epsilon$ and increasing $N$ and $l_\textrm{max}$. The simulation extended up to $R \sim 0.35$. A global fit with a linear function gives
\be\label{eq:linear1}
S_\textrm{fit}(R) = 0.4638\, \frac{R}{\epsilon}-0.06\,,
\ee
and is also shown in the figure. The excellent agreement establishes the area law in our proper distance lattice, consistent with having a Markovian cutoff. Furthermore, the value of $F$ (the constant term) is in very good agreement with the theoretical result (see e.g. \cite{Casini:2015woa}).

\begin{comment}
On the other hand, for larger $R$ we start to find small deviations from the exact linear behavior. Mostrar figuras de F(R) para m=0 y 2 o 3 cutoffs, hasta el rango de R=1. The deviation from a constant $F$ means that we are not recovering the continuum AdS isometries and the Markov property. The violation is quite subleading, as the area term is the correct one, but is nevertheless nonzero. One can check that the results converge in $\epsilon$ and $l_\textrm{max}$, so this should be a finite size effect. This is quite expected, since the conformally coupled scalar (with $m_0=0$ and hence a negative mass squared contributed by the conformal coupling) is extremely sensitive to the asymptotic boundary condition. The deviation from $F=\textrm{const}$ is most probably attributed to the fact that we are fixing a Dirichlet boundary condition at $\rho= \epsilon N$, and then taking $N$ large, while the correct procedure is to set it directly at infinity.
\end{comment}

\begin{figure}[h]
    \centering
    \includegraphics[width=0.6\linewidth]{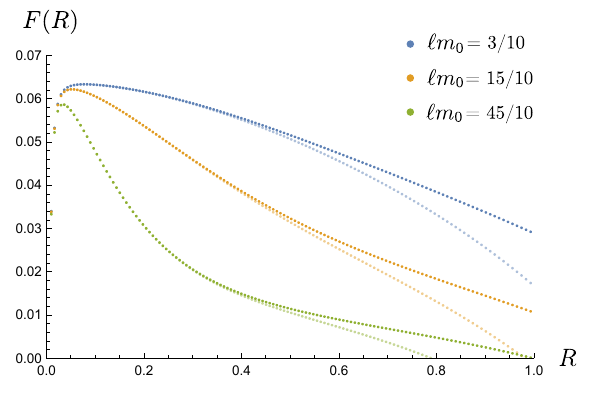}
    \caption{$F(R)$ as a function of $R$ for various values of the mass. The different cut-offs read $\epsilon=1/300$, $N=3000$. For each mass we perform the angular momentum sum up to $l_\textrm{max}=10 \,\times\,10^3$ (lighter-colored points) and  $l_\textrm{max}=15\,\times\,10^3$ (darker-colored points) in order to exhibit the convergence of the sum.}
    \label{fig:F1}
\end{figure}

Next, in Fig. \ref{fig:F1} we plot running F-functions $F(R)=R S'(R)-S(R)$ for the massive flows, calculated using a cubic interpolation. Discarding the first few points (which are strongly affected by lattice effects), we can see it  approaches the expected value $F(0)\approx 0.0638$ for short intervals \cite{Jafferis:2011zi} and it monotonically decreases as we increase the interval length. We present results for three different masses, and for each mass we show angular momentum sums up to two different $l_\textrm{max}$, in order to illustrate the convergence with this parameter. In particular, we find that obtaining accurate results for larger $R$ requires increasing $l_\textrm{max}$. This exemplifies the F-theorem in lattice RG flows.

\subsection{Entanglement in AdS$_4$}\label{subsec:Ed4}

Let us finish by presenting our results in four dimensions.
In this dimensionality, we calculate the entropic measure of degrees of freedom $\Delta A(R)=\frac12(R^2 \Delta S''(R)-R \Delta S')\le 0$. This quantity  should be negative  because of our irreversibility formula \eqref{eq:irrev_formula} and vanish in the small interval limit; $A(R)$ for the conformal fixed point should also reproduce the known A-anomaly. We stress though that the irreversibility formula does not require this quantity to be monotonic.

\begin{figure}[h]
    \centering
    \includegraphics[width=0.6\linewidth]{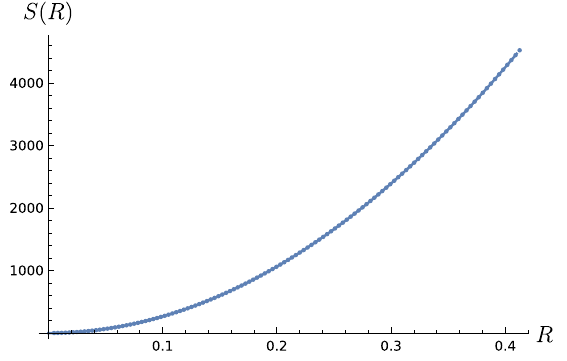}
    \caption{Lattice results (dots) for $\epsilon=1/300$, $N=3000$, $l_\textrm{max}=4\times10^4$ for the scalar with $m_0=0$, and global fit (\ref{eq:fitAglobal}).}
    \label{fig:pAm0}
\end{figure}

We have first studied the conformal fixed point with a very small cutoff, in order to check that in the continuum limit we recover the theoretical result (\ref{eq:EE4dm0}), and hence that the lattice provides a Markovian cutoff. In Fig. \ref{fig:pAm0} we present the lattice results for $\epsilon=1/300$, $N=3000$, $l_\textrm{max}=40.000$ for the scalar with $m_0=0$. We also show the global fit according to (\ref{eq:EE4dm0}), which gives
\be\label{eq:fitAglobal}
S_\textrm{fit}(R)= 0.295 \,\frac{R^2}{\epsilon^2}+0.011\,\log R-0.016\,.
\ee
The area term and log term coefficients furthermore coincide, as expected, with the CFT result in a flat space (for a lattice analysis see \cite{Lohmayer:2009sq}).
The excellent agreement shows that the lattice is providing a Markovian regulator, and that the EE for the CFT in AdS agrees with the theoretical predictions.

\begin{figure}[h]
    \centering
    \includegraphics[width=0.85\linewidth]{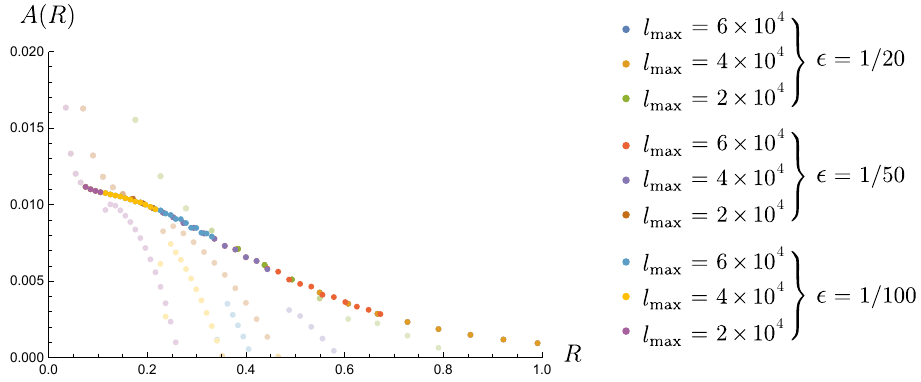}
    \caption{$A(R)$ for $m=15/10$ in units of the AdS radius, for $\epsilon=1/20,1/50$ and $1/100$, computed truncating the sum \eqref{eq:EEl} up to $l_{\max}=2\times10^4,4\times10^4$ and $6\times10^4$. In all cases, $N$ was chosen such that $\epsilon N=10$. The darker points determine a curve of converged results.}
    \label{fig:Aenvelope}
\end{figure}

Finally, in Fig. \ref{fig:Aenvelope} we plot the running A-function $A(R)=\tfrac{1}{2}(R^2S''(R)-RS'(R))$ for $m_0=15/10$, computed using different values for the cut-offs $\epsilon$ and $l_\textrm{max}$. Here we find two competing numerical limitations. On the one hand, in order to obtain accurate results for very small intervals (the ultraviolet regime), we need very small $\epsilon$. But on the other hand, obtaining accurate converged results for large $R$ requires large $\epsilon l_\textrm{max}$. It is numerically very expensive to then consider a very small $\epsilon$ throughout all the $R$ span. Instead of that, we have superposed several choices of $\epsilon$ and $l_\textrm{max}$ (checking their convergence) so that we can access larger values of $R$. The resulting converged curve is shown with dark points. These lattice results verify our entropic inequality $\Delta A(R) \le 0$; we also observe that $A(R)$ is monotonically decreasing, but so far we don't have analytic tools to establish this for more general theories.\footnote{The same monotonic behavior was found in flat space \cite{Boutivas:2025qan}.}

%%%%%%%%%%%%%%%%%%%%%%%%%%%%%%%%%%%%%%%%%
%%%%%%%%%%%%%%%%%%%%%%%%%%%%%%%%%%%%%%%%%
%%%%%%%%%%%%%%%%%%%%%%%%%%%%%%%%%%%%%%%%%
%%%%%%%%%%%%%%%%%%%%%%%%%%%%%%%%%%%%%%%%%
\section{Conclusions}\label{sec:concl}

In this paper we proved the Markov property for null entangling surface deformations for CFTs in AdS, which we then used to determine the entanglement structure. We then established the irreversibility of renormalization group flows for unitary quantum field theories living in rigid anti–de Sitter spacetime in $2\leq d \leq 4$. Our proof is based on entropic methods and exploits strong subadditivity together with AdS invariance to derive a universal differential inequality, Eq. (\ref{eq:introDS}), for spherical entangling regions. We illustrated and tested these general theorems in explicit free-field models, developing lattice formulations adapted to AdS geometry and computing entanglement entropies and RG charges both analytically and numerically. These examples clarify the structure of RG flows in AdS and highlight the distinction between conformal and massive theories in curved spacetime.

This work develops a first set of quantum information–theoretic tools for quantum field theories in rigid AdS spacetime. The results presented here open several natural directions for further investigation:

\begin{itemize}

\item We obtained analytic results for the Dirac fermion in AdS$_2$. It would be interesting to extend this to the scalar field, and to use the analytic expressions in order to derive short and long distance expressions for the C-function.

\item In this work we focused on quantum information methods to prove the irreversibility of the RG. In flat spacetime, irreversibility results can also be obtained from correlator methods (see e.g. \cite{Zamolodchikov:1986gt,komargodski2011renormalization,Hartman:2023qdn}). It would be very interesting to extend these correlator methods to AdS.

\item A natural extension is the study of strongly coupled theories using holography. In this context, RG flows in rigid AdS correspond to warped bulk geometries whose conformal boundary is an AdS$_{d-1}$ spacetime. It would be interesting to understand how the entropic inequality derived here is realized geometrically and whether it leads to new constraints on admissible holographic flows.

\item Our analysis in $d=2$ suggests that novel phenomena may arise when considering defects embedded in AdS. In this case, the boost-invariance argument underlying our proof does not directly apply, and it would be important to determine whether irreversibility can still be established using alternative information-theoretic methods.

\item In flat space, mutual information imposes nontrivial constraints on the operator content of conformal field theories. Extending such constraints to CFTs in AdS, where the geometry modifies long-distance behavior, may provide new insights into allowed spectra and correlation structures.

\item Finally, it would be valuable to extend our free-field analysis to higher-spin theories, including gauge fields and gravitons, where subtleties associated with gauge invariance and boundary conditions may lead to qualitatively new features.

\end{itemize}

\subsection*{Acknowledgements}

It is a pleasure to thank Horacio Casini, Marina Huerta for encouragement and extensive discussions at various stages of this work. We would also like to thank Lorenzo di Pietro for insightful discussions. 
N.A. is supported by a CONICET PhD fellowhip.  
I.S.L. would like to acknowledge support from the ICTP through the Associates Programme (2023-2028), by a grant from the Simons Foundation whilst on the Isaac Newton Institute Ramanujan Fellowship and the Brew Sisters program for theoretical physicists. G.T. is supported by CONICET, CNEA, and Instituto Balseiro, Universidad Nacional de Cuyo. We would like to acknowledge the Simons Foundation targeted grant to institutions that funds the Simons-Balseiro Theoretical Physics Initiative.

\bibliography{EE}{}
\bibliographystyle{utphys}

\end{document}